\documentclass[conference]{IEEEtran}
\usepackage{bm}
\usepackage{epsfig}
\usepackage{graphicx}
\usepackage{subfigure}
\usepackage{float}
\usepackage{cite}
\usepackage{url}
\usepackage{color}
\usepackage{balance}
\usepackage{mdwlist}
\usepackage{multirow}
\usepackage{threeparttable}
\usepackage{enumitem}
\usepackage{amsmath}
\usepackage{stmaryrd}
\usepackage{booktabs}
\usepackage{siunitx}
\usepackage[numbers,sort&compress]{natbib}
\usepackage{threeparttable}
\usepackage{epsfig,amsmath,amsfonts}
\usepackage{amsfonts}

\makeatletter
\newif\if@restonecol
\makeatother

\usepackage[ruled,vlined]{algorithm2e}
\usepackage{algpseudocode}
\usepackage{amsmath}

\newcommand{\parm}{{\xi}}
\newcommand{\vecpar}{\boldsymbol{\parm}}
\newcommand{\matpar}{\boldsymbol{\bar \parm}}
\newcommand{\weight}{w}
\newcommand{\vecweight}{\boldsymbol{\weight}}
\newcommand{\vece}{\mat {e}_1}

\newcommand{\multiGPC}{\Psi }

\newcommand{\polyInd}{\alpha}
\newcommand{\basisInd}{\boldsymbol{\polyInd}}

\newcommand{\ten}[1]{\mathcal{#1}}

\newcommand{\mat}[1]{\mathbf{#1}}

\newcommand{\reff}[1]{(\ref{#1})}

\newcommand{\hzc}[1]{\textcolor{black}{#1}}

\RequirePackage{url}

\def\BibTeX{{\rm B\kern-.05em{\sc i\kern-.025em b}\kern-.08em
    T\kern-.1667em\lower.7ex\hbox{E}\kern-.125emX}}
\begin{document}

\title{Efficient Uncertainty Modeling for System Design via Mixed Integer Programming}


\author{\IEEEauthorblockN{Zichang He$^\star$, Weilong Cui$^\dagger$, Chunfeng Cui$^\star$, Timothy Sherwood$^\dagger$ and Zheng Zhang$^\star$}
\IEEEauthorblockA{{$^\star$Department of Electrical and Computer Engineering,\; \;$^\dagger$Department of Computer Science} \\
{University of California, Santa Barbara, CA 93106}\\
Emails: zichanghe@ucsb.edu, cuiwl@cs.ucsb.edu, chunfengcui@ucsb.edu, sherwood@cs.ucsb.edu, zhengzhang@ece.ucsb.edu}
}

\maketitle
\begin{abstract}
The post-Moore era casts a shadow of uncertainty on many aspects of computer system design.  Managing that uncertainty requires new algorithmic tools to make quantitative assessments. While prior uncertainty quantification methods, such as generalized polynomial chaos (gPC), show how to work precisely under the uncertainty inherent to physical devices, these approaches focus solely on variables from a continuous domain.  However, as one moves up the system stack to the architecture level many parameters are constrained to a discrete (integer) domain.
This paper proposes an efficient and accurate uncertainty modeling technique, named mixed generalized polynomial chaos (M-gPC), for architectural uncertainty analysis. The M-gPC technique extends the generalized polynomial chaos (gPC) theory originally developed in the uncertainty quantification community, such that it can efficiently handle the mixed-type (i.e., both continuous and discrete) uncertainties in computer architecture design.
Specifically, we employ some stochastic basis functions to capture the architecture-level impact caused by uncertain parameters in a simulator. 
We also develop a novel mixed-integer programming method to select a small number of uncertain parameter samples for detailed simulations. 
With a few highly informative simulation samples, an accurate surrogate model is constructed in place of cycle-level simulators for various architectural uncertainty analysis. 
In the chip-multiprocessor (CMP) model, we are able to estimate the propagated uncertainties with only 95 samples whereas Monte Carlo requires $5\times {10^4}$ samples to achieve the similar accuracy.
We also demonstrate the efficiency and effectiveness of our method on a detailed DRAM subsystem.
\end{abstract}


\section{Introduction}
Designing a computer system in an era of rapidly evolving applications and technology nodes involves many uncertainties.  Computer system design have been known to be susceptible to all sorts of uncertainties from device-level process variations~\cite{Mittal:2016:SAT:2891449.2871167} to variations in application characteristics inside a datacenter~\cite{Mishra:2010:TCC:1773394.1773400}.
Of course there is a deep library of work on quantifying uncertainty in architecture and system design that has been particularly focused on device and circuit level uncertainty~\cite{Borkar:2003:PVI:775832.775920,Bhardwaj:2005:LMN:1065579.1065719,Zhang:2009:PVC:1629911.1630092, Wong:2005:FDA:1129601.1129608, das2007mitigating,7271059Yan} for us to draw upon, but as one moves up the system stack from the device to the architecture level and above many variables (e.g., cycles to satisfy an L1 cache miss or the number of bits of error correcting to use) are constrained to a discrete (e.g., integer) domain.



Quantifying uncertainty at the system level has been demonstrated via some high-level analytic models~\cite{Cui:2017:EUA:3123939.3124541}, yet doing so with detailed simulation is extremely challenging. On one hand, Monte Carlo (MC) methods require a large amount of data samples of the performance outputs due to its slow convergence rate. The time cost associated with acquiring such samples renders these techniques unsuitable when detailed simulator is used instead of analytic models, \hzc{where one simulation sample usually costs minutes or hours (or even up to days in some large-scale system simulations)}. On the other hand, many advanced uncertainty quantification methods such as generalized polynomial chaos (gPC), although highly efficient, cannot handle the unique challenge of such a mixed-domain (continuous and discrete) problem.

We demonstrate a new algorithm for accurate uncertainty analysis in the context of computer system design by using only a small number of detailed simulations. To achieve this goal, we extend the generalized polynomial chaos (gPC) method~\cite{xiu2010numerical}, which is a powerful technique developed in the uncertainty quantification community. Due to its superior convergence rate and orders-of-magnitude speedup over MC in many applications, the gPC technique has been successfully applied to electronic design automation problems. 
Existing work includes fast stochastic modeling, simulation and optimization for electronic integrated circuits~\cite{zhang2013stochastic,zhang2013uncertainty,zhang2016big}, integrated photonic devices and circuits~\cite{cui2018uncertainty},  micro-electromechanical systems~\cite{agarwal2009domain} and so forth. However, all these techniques can only handle analog behaviors/performance and only consider continuous uncertain parameters.


{\bf Paper Contributions.} In this paper, we develop a new surrogate modeling technique, named mixed-gPC (M-gPC), for accurate uncertainty analysis of computer system design. 
Our algorithm employs a mixed integer programming method to handle mixed continuous and discrete uncertain parameters with a small number of simulations. To verify this theory experimentally in the context of computer system design, we examine its application to both a closed-form chip-multiprocessor (CMP) model and a detailed simulation-driven DRAM subsystem. Our specific contributions include:
\begin{itemize}
    \item We present a CAD framework for architectural uncertainty analysis. Based on the given continuous and discrete distributions of input uncertain parameters, our algorithm automatically chooses the basis functions and simulation samples to build an accurate surrogate model. 
    
    \item We present the numerical algorithms of building basis functions and choosing simulation samples. The key challenge is to decide a small number of highly informative simulation samples required in the stochastic collocation framework~\cite{xiu2005high}. We formulate this problem as a mixed-integer programming (MIP), and present a hierarchical decomposition method to efficiently solve it. 
    
    \item We validate the proposed M-gPC framework by an analytical CMP model and a realistic DRAM subsystem. Our framework shows significant speedup and high accuracy on these two architectural analysis benchmarks.
\end{itemize}

\section{Architectural Uncertainty Analysis Framework}
\label{motivation}


We propose a M-gPC modeling method for computer architectural uncertainty analysis.
The basic idea is to use a mixed-integer optimization method to select a few important samples for both continuous and discrete uncertain parameters. Then, an accurate stochastic surrogate model is constructed for architecture-level uncertainty analysis after a small number of detailed cycle-level simulations. 
\begin{figure}[t]
\centering
\includegraphics[width=\linewidth]{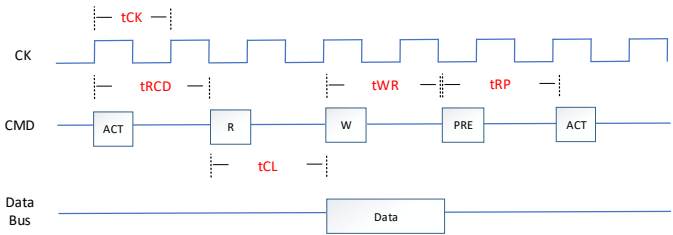}
\caption{All named DRAM timings here are considered under the influence of uncertainty (process variation). Note that the timings are not proportional to real number of cycles in this figure.}\label{fig:dram}
\end{figure}

\subsection{Problem Formulation and Overall Workflow}

{\bf Problem Formulation.} Given a computer architecture design case (e.g., a DRAM subsystem, see Fig.~\ref{fig:dram}), let $\mat{y}(\vecpar)$ be some uncertain performance metrics of interest, such as the bandwidth or power. 
We use vector $\vecpar = \left[ {{\parm_1},{\parm_2}, \ldots ,{\parm_d}} \right]^T \in {\mathbb{R}^d}$ to denote some model uncertainty parameters, such as the time of one tick of clock (tCK), time precharge/recovery period (tRP), and so forth. 
We assume that these parameters are mutually independent, and each parameter $\xi_i$ admits a probability density function $\rho_i(\parm_i)$.
Please note that these uncertain parameters can be {\it either continuous or discrete}. 
For instance, tCK often obeys a truncated Gaussian distribution~\cite{sarangi2008varius}, and tRP often follows a binomial distribution~\cite{chandrasekar2014exploiting} due to the process variations. Our goal is to quantify the uncertainty of $\mat{y}(\vecpar)$ caused by $\vecpar$. Because each detailed cycle accurate simulation is extremely expensive, we instead want to build a surrogate model in the following form with only {\it a small number of simulations}:
\begin{equation}
\label{eq:expansion}
    \mat{y}(\vecpar)\approx \sum_{|\basisInd|=0}^p \mat{c}_{\basisInd}\multiGPC_{\basisInd}(\vecpar).
\end{equation}
Here, $\multiGPC_{\basisInd}(\vecpar)$ is a stochastic basis function indexed by vector $\basisInd$, and $p$ upper bounds the total order of the basis functions.

\begin{figure}[t]
\centering
\includegraphics[width=0.9\linewidth]{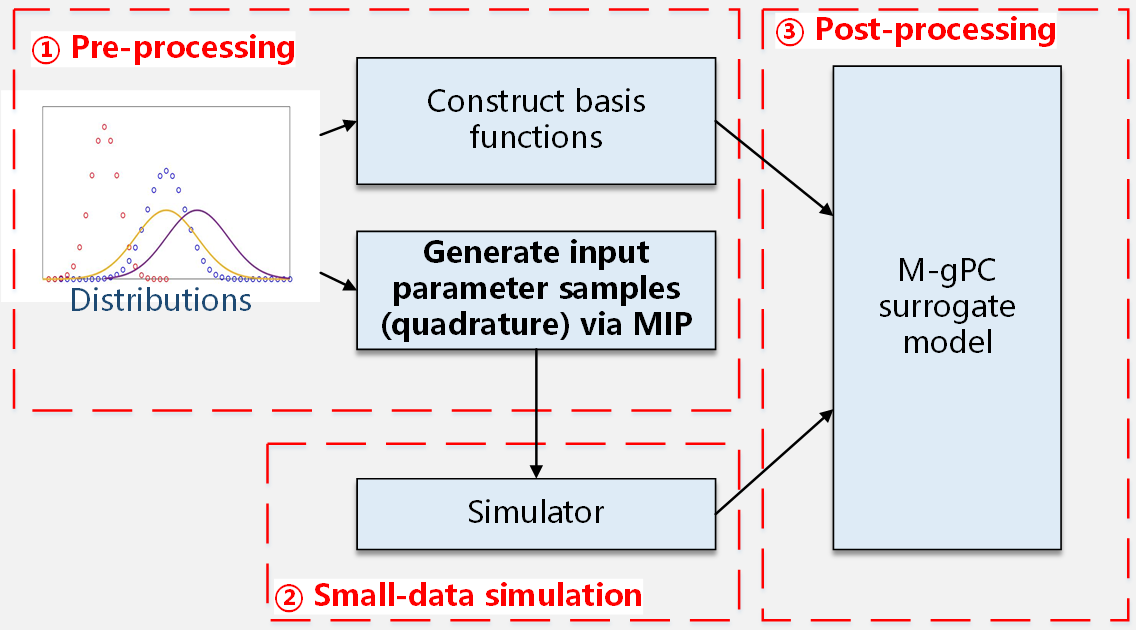}
\caption{Workflow of the architectural uncertainty analysis method.}\label{fig:workflow}
\end{figure}

\vspace{5pt}
{\bf Overall Workflow.} Before proceeding to the technical details, we first describe our high-level workflow in Fig.~\ref{fig:workflow}.
\begin{itemize}
    \item {\bf Step 1: pre-processing.} Based on the given distributions of uncertain parameters $\vecpar$ (which can be either continuous or discrete), we need to specify a class of basis functions $\{ \multiGPC_{\basisInd}(\vecpar)\}$. We also decide a few highly informative sample points and weights $\{ \vecpar_i, w_i\}_{i=1}^M$. Here $w_i$ quantitatively describes the importance of sample $\vecpar_i$.
    
\item {\bf Step 2: small-data simulations.} Given the carefully selected parameter samples, we call a high-fidelity cycle accurate simulator (e.g., the DRAMSim2~\cite{rosenfeld2011dramsim2} for a DRAM) to simulate sample $\vecpar_i$ for $i=1,2\dots, M$. The repeated simulations produce a set of output performance samples $\{ \mat{y} (\vecpar_i)\}_{i=1}^M$. Because $M$ is very small, these detailed simulations can be done in a relatively short time.

\item {\bf Step 3: post-processing.} With $\{ \mat{y} (\vecpar_i)\}_{i=1}^M$ and the importance of each sample, we will decide the weight vector $\mat{c}_{\basisInd}$ for each basis function in Eq.~\eqref{eq:expansion}. 

\end{itemize}
\begin{figure*} 
    \centering
    \includegraphics[width=6.2in]{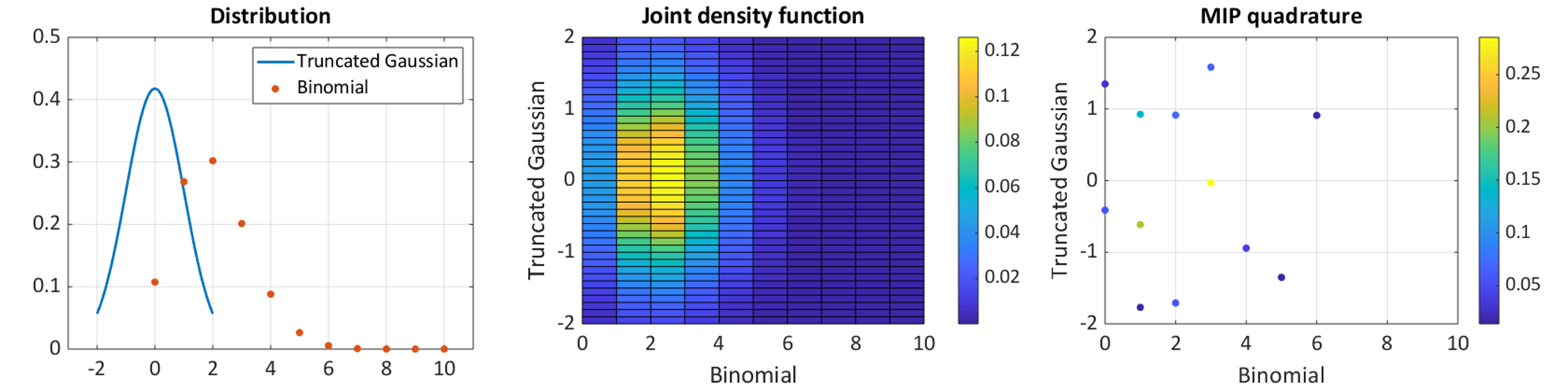}
    \caption{Example of mixed-integer programming-based quadrature for a discrete variable with distribution $\text{Binomial}(10,0.2)$ and another continuous one with a truncated Gaussian distribution in the range $(-2,2)$. The color bar of MIP quadrature represents the weights of all samples.}
    \label{fig:Qpoints}
\end{figure*}

\subsection{The Proposed M-gPC Method}
Three questions need to be answered in the above workflow:
\begin{itemize}
    \item {\bf Q1:} How shall we decide the basis functions?
    \item {\bf Q2:} How shall we decide the parameter sample $\vecpar_i$ and  weight $w_i$, and the proper number of samples $M$?
    \item {\bf Q3:} How shall we post-process the data to get the surrogate model \eqref{eq:expansion}?
\end{itemize}

The first and third questions are already addressed in the standard gPC method~\cite{xiu2010numerical}. Given the probability density function of each parameter, the gPC method selects a set of orthornormal polynomials as the basis functions:
\begin{equation}
 \mathbb{E}\left[{\multiGPC}_{\basisInd} \left( \vecpar \right) \multiGPC_{\boldsymbol{\beta }}\left( \vecpar \right)\right ]=
\left\{\begin{array}{cc}
1,&{\text{if}} \; \basisInd=\boldsymbol{\beta };\\
0,&\text{otherwise.}
\end{array}\right.
\end{equation}
Here the index vector $\basisInd=[\alpha_1,\ldots,\alpha_d]^T$ denotes the polynomial degree of each uncertain parameter in the basis function. As a result, the expansion \eqref{eq:expansion} employs $N_p=\binom{d+p}{p}$ basis functions in total. This choice of basis functions allows the stochastic collocation method~\cite{xiu2010numerical} to compute each unknwon $\mat{c}_{\basisInd}$ by a projection procedure:
\begin{equation}
\label{eq:pseudo_proj}
\mat{c}_{\basisInd}= \mathbb{E}\left[ \mat{y} \left( {\vecpar} \right) {\Psi_{\basisInd}} \left( {\vecpar} \right) \right]  \approx \sum\limits_{i = 1}^M { \mat{y} \left( \vecpar_i \right) \Psi_{\basisInd}\left( \vecpar_i \right) w_i }.
\end{equation}
Here $\{ \vecpar_i, w_i\}_{i=1}^M$ are the quadrature samples and weights that we need to determine. Once the surrogate model is constructed, we can easily obtain the mean value and standard deviation of the output performance via Eq.~\reff{eq:moments}: 
\begin{equation}\label{eq:moments}
    \mathbb{E}[\mat{y}(\vecpar)]\approx \mat{c}_{\mathbf 0},\quad \sigma[\mat{y}(\vecpar)]\approx \sqrt{\sum_{|\basisInd|=1}^p \mat{c}_{\basisInd}^2},
\end{equation}
where all calculations are performed component-wise. We can also obtain the histogram or probability distribution of $\mat{y}(\vecpar)$ without calling the detailed cycle accurate simulator again. The details of basis functions are given in Appendix~\ref{sec:TTR}.

\vspace{5pt}
{\bf Sample and weight selection in M-gPC.} However, addressing the second question is difficult. According to \eqref{eq:pseudo_proj}, we need to choose a small number of samples and weights to accurately estimate the numerical integration. Numerous numerical integration rules (e.g., Gauss quadrature rule and sparse grids~\cite{xiu2010numerical}) are available for continuous variables, but they cannot handle discrete variables. A naive choice is to use all possible samples of a discrete uncertain variable, but this leads to a huge number of simulation samples. In order to address this issue, we develop a novel mixed-integer programming method to generate a few high-quality quadrature samples and weights. We defer the details of this approach to Section~\ref{sec:proposed_method}, and we first demonstrate this approach by a simple example. 
\hzc{We consider a problem with two uncertain variables: the continuous one is truncated Gaussian in the range of $(-2,2)$, and the discrete one obeys a binomial distribution.} When choosing quadrature samples, the first variable can be any value in the range of $(-2,2)$, but the second one can only take some integer values below 10. The results of our mixed integer programming are shown in Fig.~\ref{fig:Qpoints}: only 12 samples with different weights are generated.

\section{Proposed MIP-Based Quadrature Rule}
\label{sec:proposed_method}

This section presents our mixed-integer programming solver to determine the quadrature points and weights $\{\vecpar_i, w_i\}_{i=1}^M$ required in Eq. \reff{eq:pseudo_proj}. The quadrature samples are provided to a cycle-level simulator for repeated simulations. Therefore, we hope to make $M$ as small as possible.


\subsection{Optimization Formulation}

Let $\mathcal{I}=\{i \,|\, \xi_i\in\mathbb{Z}\}$ denote the index set of integer parameters in $\vecpar$. We propose to compute the quadrature points and weights by an optimization-based method.  Our method differs from~\cite{cui2018stochastic,cui2018stochastic_journal}: our proposed method can handle both discrete and continuous uncertain parameters, whereas~\cite{cui2018stochastic,cui2018stochastic_journal} can only handle continuous uncertain variables.

Eq.~\reff{eq:pseudo_proj} requires the integration of polynomials up to order $2p$ if $y(\vecpar)$ can be well approximated by an order-$p$ expansion. Therefore, our optimization problem is set up by matching the integration of basis functions up to order $2p$. If $| \basisInd | \leq 2p$, then there is a one-to-one correspondence between $\basisInd$ and integer $k\in [1, N_{2p}]$ where $N_{2p}=(d+2p)!/(d!)(2p!)$. We use the scalar index for simplicity, and seek for $\{\vecpar_i,w_i\}_{i=1}^M$ such that 
\begin{align}
  \nonumber  \mathbb{E}[\multiGPC_k(\vecpar)]= \sum\limits_{i = 1}^M \Psi_k \left( \vecpar_i \right)w_i, \ \forall\, k\in[N_{2p}]\\
   \text{where } w_i\ge0\text{ and }\vecpar_{i,\ten I}\in\mathbb{Z}^{|\ten I|}. 
    \label{expectation_basis}
\end{align}
Here, $\vecpar_{i,\ten I}=\{\vecpar_{i,l}\}_{l\in\ten I}$ denotes all integer elements in sample $\vecpar_i$.  
Due to the orthonormal condition, we know $\mathbb{E}\left[ {{\Psi _k}\left( {\vecpar} \right)} \right] = \mathbb E\left[ {{\Psi _k}\left( {\vecpar} \right){\Psi _1}\left( {\vecpar} \right)} \right] = {\delta _{1k}}$.

As a result, we  solve $\{\vecpar_i,w_i\}_{i=1}^M$ via the following nonlinear least squares 
\begin{align}
    \min_{\bar{\vecpar},\mat w}\quad  \|\boldsymbol{\Phi}(\bar \vecpar)\mat w-\vece\|_2^2,\quad
    \text{s.t.}\quad \mat w\ge 0,\ \bar \vecpar_{\ten I}\in\mathbb{Z}^{M|\ten I|}.
   \label{optimization_formula} 
\end{align}
where $\matpar = {\left[ {{\vecpar}_1,{\vecpar}_2, \ldots ,{\vecpar}_M} \right]^T} \in {\mathbb{R}^{M\times d}}$, $\bar\vecpar_{\ten I}=\{\vecpar_{i,\ten I}\}_{i=1}^M$, 
$\vecweight = {\left[ {{w_1},{w_2}, \ldots ,{w_M}} \right]^T} \in {\mathbb{R}^M}$,
$\vece = {\left[ {1,0, \ldots ,0} \right]^T} \in {\mathbb{R}^{{N_{2p}}}}$, 
$\boldsymbol \Phi \left( \matpar \right) \in {\mathbb{R}^{{N_{2p}} \times M}}$ with each elements ${\left[ {\boldsymbol\Phi \left( \matpar \right)} \right]_{ki}} = {\Psi _k}\left( {{{\vecpar}_i}} \right)$. 
There are ${M \times \left( {d + 1} \right)}$ unknown variables in total, which can be a large-scale optimization problem as $d$ increases. 

\subsection{Hierarchical BCD for Solving Problem \reff{optimization_formula}}

Eq. \reff{optimization_formula} is a large-scale mixed integer nonlinear programming problem (MINLP) with $M|\ten I|$ integers. We solve this hard problem by a hierarchical decomposition method. 
Our key idea is to simplify the original large scale MINLP into several easier sub-problems. 

We intend to employ a block coordinate descent (BCD) method to solve the quadrature points and weights separately.
However, directly applying it to our mixed-integer case is inefficient since the complexity of mixed-integer programming grows exponentially with the number of integer variables~\cite{grossmann2002review}.
In order to further speed up the algorithm,  we  separate the quadrature points into $M$ blocks. Consequently, we only need to solve a MINLP with $|\ten I|$ integer variables for each sub-problem. Our detailed framework is shown in Alg.~\ref{Solve_optimization_formula}.  
\begin{algorithm}[t]
  \caption{Hierarchical BCD for solving \reff{optimization_formula}}
  \label{Solve_optimization_formula}
  \KwIn{Initial points and weights, maximal outer iteration ${n_{max}}$, and solver error tolerance $\varepsilon $.}
  \KwOut{Optimized points and weights $\left\{ {{{\vecpar}_i},{w_i}} \right\}_{i = 1}^M$;}
  \For{${t}= 1,2, \ldots ,{n_{\max }}$}
  {
  \For{$i =1,2, \ldots ,M$}{
   Solve $\Delta {{\vecpar}_i}$ via Eq. \reff{regularization_opt};\\
   Update $i$th point via solving Eq. \reff{update_point};\\
   Update all weights via solving Eq. \reff{sub_weight};\\
   }
   \If{$\left\| {\Phi \left( {{\matpar^{{t}}}} \right){{\vecweight}^{{t}}} - \vece} \right\|_2^2 \le \varepsilon$}
    {
      break; \% converge
     }
    \Else
    {
    \If{$\left\| {\Delta \matpar} \right\|_F \le {10^{-8}} $} 
    {break; \% not converge}
    }
  }
\end{algorithm}

In Alg.~\ref{Solve_optimization_formula}, we have an outer iteration ${t}$ and an inner iteration $i$. 
In this case, only one sample is optimized in each inner iteration, which only has $d$ unknown variables. 
By applying the Gauss Newton method, the original problem is converted to a mixed-integer quadrature program (MIQP), which is much easier to solve.
Considering that the Jacobian matrix $J$ can be ill-conditioned, we adopt the Tikhonov regularization~\cite{hansen1992analysis} here to make the MIP solver more stable:
\begin{align}
 \nonumber  \min\limits_{\Delta \vecpar_i}\quad & {\left\| {\hat{\mat J}_i^{{t}\times i}\Delta {{\vecpar}_i} + {\mat r^{{t}\times i}}} \right\|_2^2 + {\lambda}\left\| {\Delta {{\vecpar}_i}} \right\|_2^2}\\ 
   \text{s.t.}\quad & \Delta \vecpar_{i,\ten I}\in\mathbb{Z}^{|\ten I|},
  \label{regularization_opt}
\end{align}
where $\mat r^{{t}\times i} = \boldsymbol\Phi \left( {{\matpar^{{t}\times i - 1}}} \right){{\vecweight}^{{t}\times i}} - \vece$ denotes the residual given the $i$th points under the $\left({t} \times i\right)$th iteration and ${\mat J_i^{{t}\times i}}$ denotes the Jacobian matrix of ${\mat r^{{t}\times i}}$, and $\lambda$ is a regularization parameter.
Then we can update each quadrature point as
\begin{equation}\label{update_point}
{\vecpar}_i^{t} = {\vecpar}_i^{t-1} + \Delta {{\vecpar}_i}.
\end{equation}
After one sample is optimized, we fix all the points and update the weights via solving a linear least square problem:
\begin{equation}\label{sub_weight}
{{\vecweight}^{{t} \times i}} = \arg \mathop {\min }\limits_{\vecweight} \left\| {\boldsymbol\Phi \left( {{\matpar^{{t} \times i}}} \right)\vecweight - \vece} \right\|_2^2
\end{equation}
 
In summary, we decompose the original large-scale MINLP problem [Eq. \reff{optimization_formula}] to many linear least square problems [Eq. \reff{sub_weight}] and small-scale MIQP problems [Eq. \reff{regularization_opt}], which are much more efficient to solve. The comprehensive algorithm is summarized in Alg. \ref{Solve_optimization_formula}. With the proposed MIP-based quadrature, we can select a small number of samples to calculate the M-gPC coefficients via Eq.~\reff{eq:pseudo_proj}. 

\subsection{Initialization and Number of Quadrature Points}
\label{sec:initialization}
  \begin{algorithm}[t]
  \caption{MIP-based stochastic collocation.}
  \label{SC_algorithm}
  \begin{algorithmic}[1]
    \Ensure
      The M-gPC coefficients $\left\{ {\mat{c}_{\basisInd}} \right\}_{|\basisInd|=0}^{p}$
    \State Initialize the quadrature points and weights;
    \label{Alg_initial}
    \State \textbf{Increase Phase.} Update points and weights via Alg. \ref{Solve_optimization_formula}. If algorithm fails to converge, increase the number of points and go  back to last line.
    \State \textbf{Decrease Phase.} Update points and weights via Alg. \ref{Solve_optimization_formula}. If algorithm converges, decrease the number of points until it fails.
    \State Call the simulator to obtain $\left\{ {{\mat y(\vecpar_i)}} \right\}_{i = 1}^M$
    \State Calculate the M-gPC weight vectors via Eq. \reff{eq:pseudo_proj}.
  \end{algorithmic}
\end{algorithm}

In practice, a global optimal solution to problem \reff{optimization_formula} is unnecessary: any solution with a high accuracy and leading to a small number of quadrature points
\hzc{can be used to build a M-gPC surrogate model with good performance.}
However, choosing a good initial guess is important to ensure high accuracy. We employ the weighted complete linkage clustering method~\cite{cui2018stochastic} to generate the initial guess. 
Firstly, many candidate samples with weights are randomly generated by Monte Carlo. Then, they are clustered to a smaller number of points. In this process, any two points with a minimal weighted distance can be merged into one point sequentially until the number of points achieves the initial setting. 
The weighted distance between two sample points is defined as follows:
\begin{equation}
{D_{ij}} = \left( {{w_i} + {w_j}} \right)\left( {\mathop {\max }\limits_{{\vecpar _1} \in {C_1},{\vecpar _2} \in {C_2}} d\left( {{\vecpar _1},{\vecpar _2}} \right)} \right).
\end{equation}
In order to properly determine the number of quadrature points $M$, an increase-decrease module~\cite{cui2018stochastic_journal} is adopted. 
Firstly, we increase $M$ to ensure that \reff{optimization_formula} can be solved with a high accuracy. Then, we decrease $M$ and solve \reff{optimization_formula} to check if an accurate quadrature rule can be found with fewer samples.
Integrating this module with our framework, the overall MIP-based stochastic collocation method is summarized in Alg. \ref{SC_algorithm}.

We have the following remarks. \hzc{Firstly, the CPU time of the current hierarchical algorithm is negligible compared to the CPU time cost by sample simulations, especially for low-dimensional problems.}
Secondly, the theoretical guarantees on the approximation error and on the number of quadrature points in~\cite{cui2018stochastic_journal} still hold in our proposed MIP-based framework.   


\section{Case Studies}
\label{numerical_results}
\subsection{Analytical CMP Models}\label{sec:CMP}
\subsubsection{Uncertainties in Analytical CMP Models}
Here, an analytical CMP model in~\cite{Cui:2017:EUA:3123939.3124541} is used to verify the proposed architectural uncertainty analysis framework. 
A more general heterogeneous core selection problem based on that of Hill and Marty~\cite{hill2008amdahl} is illustrated in Table \ref{syn_closed_form}.
\begin{table}[!h]
\caption{Closed form of CMP model.}
\label{syn_closed_form}
\centering
\begin{tabular}{c}
\toprule
$\text{Speedup} = {1}/ \left( T_{\text{sequential}} + T_{\text{parallel}} \right)$ \\ 
${T_{\text{sequential}}} = \left(1 - f + c \times  \sum\limits_{i \in \text{core~types}} {N_{\text{core}_i}} \right) /   {P_\text{{serial}}}$\\
${T_\text{{parallel}}} = {f} / {P_{\text{parallel}}} $\\
$P_\text{serial} = \max \left\{ P_{\text{core}_i}  | N_{\text{core}_i} > 0 \right\}$\\
$P_\text{parallel} = \sum\limits_{i \in \text{core~types}} {N_{\text{core}_i}} \times {P_{{core}_i}} $\\
$P_{\text{core}_i} = \sqrt { A_{\text{core}_i}}$\\
$A_\text{total} = \sum\limits_{i \in \text{core~types}} {N_{\text{core}_i} \times {A_{\text{core}_i}  }}     $\\
\bottomrule
\end{tabular}
\end{table}
In terms of uncertainty description, the inputs parallelism of the application ($f$), communication overhead among cores ($c$) and designed number of each core on chip ($N_{\text{core}_i}$) are discrete, while  performance of each type of core (${P_{\text{core}_i} }$) and ${yield}_{\text{core}_i}$ are continuous. \hzc{We use the same uncertainty models described in previous work~\cite{Cui:2017:EUA:3123939.3124541} shown in Table~\ref{Tb:syn_para_set}.
Its parameter setting in a case of two cores is shown in Table \ref{Tb:syn_para_set}}.

\begin{table}[!h]
\caption{Uncertain parameters setting in the CMP model.}
\label{Tb:syn_para_set}
\centering
\begin{tabular}{|c|c|c|}
\hline
Uncertainty model &  \multicolumn{2}{c|}{Parameter setting}\\\hline
$f \sim \frac{{\text{Binomial}\left( {M,p} \right)}}{M}$                  & \multicolumn{2}{c|}{$M = 60$, $p=0.6$}  \\ \hline
$c \sim \frac{{\text{Binomial}\left( {M,p} \right)}}{M}$                  & \multicolumn{2}{c|}{$M = 80$, $p=0.7$}  \\ \hline
\multirow{2}{*}{${N_{\text{core}_i} } \sim \text{Binomial}\left( {M,{yield}_{\text{core}_i}  } \right)$} & \multicolumn{2}{c|}{$M=20$, ${yield}_{\text{core}_0}=0.7432$} \\ \cline{2-3} 
                   & \multicolumn{2}{c|}{$M=20$, ${yield}_{\text{core}_1}=0.5739$} \\ \hline
\multirow{2}{*}{${P_{\text{core}_i}  } \sim \text{Truncated Gaussian}\left( {\mu ,\sigma ,0} \right)$} & \multicolumn{2}{c|}{$\mu_0=5.6569$, $\sigma_0=1.1314$} \\ \cline{2-3} 
                   & \multicolumn{2}{c|}{$\mu_1=8$, $\sigma_1=1.6$} \\ \hline
${yield}_{\text{core}_i} = {\left( {1 + \frac{{d \times {A_{\text{core}_i} } } } {\alpha }} \right)^{ - \alpha }}$                  & $A_{\text{core}_0}=32$      &$A_{\text{core}_1}=64$     \\ \hline
\end{tabular}
\end{table}
Under such uncertain parameters, our task is to approximate the moments and distribution of Speedup as close as possible to estimate the propagated uncertainty in the model. 

\subsubsection{Numerical Results}
\begin{figure}[t]
\centering
\includegraphics[width=1\linewidth]{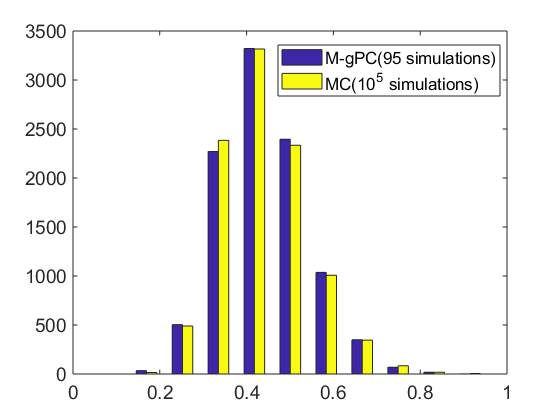}
\caption{Histograms of speedup performance with $\varepsilon = 10^{-5}$.}\label{fig:syn_error}
\end{figure}

The results of a 2nd-order M-gPC model under different thresholds $\varepsilon$ are shown in Table \ref{Tb:syn_results}. 
\hzc{A higher order leads to more accurate approximation, but order 2 already works well in this case.}
We can find that $\varepsilon$ can control the number of M-gPC samples and accuracy of the performance. The results show that the algorithm performance may be limited by a too large or too small $\varepsilon$, but it performs well with small samples in capturing the mean value and standard deviation when $\varepsilon$ ranges from $10^{-3}$ to $10^{-6}$.
In these cases, compared with the relative root mean square error (RMSE) and relative mean absolute error (MAE) between the M-gPC and MC are all around 0.03 and 0.02 respectively. 
Especially, to achieve the same level of accuracy, the sample number needed in a M-gPC model is much smaller than MC methods.
For example, if we take the results of $10^5$ MC as the ground truth, then 95 M-gPC simulation samples can already achieve the same accuracy of $5\times 10^4$ MC samples in capturing the mean value 
The histograms of $\varepsilon = {10^{-5}}$ case under $10^5$ MC samples are illustrated in Fig. \ref{fig:syn_error}.

\begin{table}[!h]
\small
\centering
\caption{Speedup performance of different models.}
\label{Tb:syn_results}
\begin{tabular}{ccccccc}
\toprule
& Sample & Mean   & Std    & RMSE  & MAE   & $\varepsilon$         \\
\midrule
\multirow{8}{*}{M-gPC} 
& 84     & 0.4353 & 0.1021 & 0.0418  & 0.0311   & 1e-2              \\%
& 85     & 0.4382 & 0.0974 & 0.0338  & 0.0231   & 1e-3        \\%
& 87     & 0.4380 & 0.0992 & 0.0306  & 0.0208   & 1e-4        \\%
& 95     & 0.4376 & 0.0986 &0.0306   & 0.0228   & 1e-5        \\%
& 123    & 0.4376 & 0.0987 &0.0314   & 0.0233   & 1e-6        \\%
& 179    & 0.4386 & 0.0982 &0.0289   & 0.0205   & 1e-7        \\%
& 182    & 0.4387 & 0.0975 & 0.0294  & 0.0214   & 1e-8        \\%
\hline
\multirow{6}{*}{MC}  & 1e3    & 0.4369 & 0.1011 & \multirow{6}{*}{N/A} & \multirow{6}{*}{N/A} & \multirow{6}{*}{N/A}\\
& 5e3    & 0.4370 & 0.1002 &      &   \\
& 1e4    & 0.4383 & 0.0995 &      &   \\
& 5e4    & 0.4375 & 0.099  &      &   \\
& 1e5    & 0.4377 & 0.0987 &      &   \\             
\bottomrule
\end{tabular}
\end{table}

\subsection{Uncertainties in Detailed Memory Subsystems}

\subsubsection{Experiment Setup}

To faithfully capture the uncertainties in a DRAM subsystem, we use the detailed DRAM simulator DRAMSim2~\cite{rosenfeld2011dramsim2}. 
We model the uncertainty parameter after previous works~\cite{sarangi2008varius,chandrasekar2014exploiting} as shown in Table \ref{Tb:DRAM_para_set} (also see Fig.~\ref{fig:dram}).
We run a collection of memory traces from SPEC 2017 benchmark suit~\cite{spec2017} shown in Table~\ref{tab:workloads}.

\begin{table}[!h]
\caption{Uncertain parameters setting in the DRAM subsystem.}
\label{Tb:DRAM_para_set}
\centering
\begin{tabular}{c}
\toprule
$\text{tCK} \sim \text{Truncated Gaussian}\left( {\mu ,\sigma ,0} \right)$\\
$\text{tRCD} \sim \text{Binomial}\left( {M,p} \right)$\\
$\text{tCL} \sim \text{Binomial}\left( {M,p} \right)$\\
$\text{tRP} \sim \text{Binomial}\left( {M,p} \right)$\\
$\text{tWR} \sim \text{Binomial}\left( {M,p} \right)$\\
\bottomrule
\end{tabular}
\end{table}
\begin{table}[!h]
\caption{Workloads and \hzc{simulation time comparison}.}
\label{tab:workloads}
\centering
\begin{tabular}{cccc}
\toprule
Workload      & Length of trace & \hzc{ MC time} & \hzc{M-gPC time}\\
\midrule
600.peribench & 46.8M           & $\sim$15.1h    &  $\sim$4.5h + 12.5m   \\
602.gcc       & 35.7M           & $\sim$8.8h     & $\sim$2.6h + 12.5m    \\
605.mcf       & 43.5M           & $\sim$14.7h    & $\sim$4.4h + 12.5m \\
623.xalancbmk & 42.9M           & $\sim$12.6h    & $\sim$3.8h + 12.5m \\
625.x264      & 30.5M           & $\sim$9h    & $\sim$2.7h + 12.5m \\
631.deepsjeng & 37.6M           & $\sim$12.3h    & $\sim$3.7h + 12.5m \\
641.leela     & 36.1M           & $\sim$11.6h     & $\sim$3.5h + 12.5m \\
998.specrand  & 32.9M           & $\sim$10.6h    & $\sim$3.2h + 12.5m \\
\bottomrule
\end{tabular}
\end{table}
 We also simulate a memory system with 1 memory channel under the JEDEC standards with a 32-entry command queue and a 32-entry transaction queue. For the DRAM itself, we experiment on a collection of different DDR3 devices shown in Table~\ref{tab:devices}. 
\begin{table}[!h]
\caption{DDR3 devices used in the DRAM subsystem.}
\label{tab:devices}
\centering
\begin{tabular}{cccc}
\toprule
Label      & Width &Capacity &Internal frequency\\
\midrule
1  & 4b &64MB &667MHz  \\
2  & 8b &32MB &400MHz \\
3  & 8b &32MB &667MHz \\
4  & 4b &32MB &800MHz \\
5  & 4b &32MB &667MHz  \\
6  & 8b &16MB &667MHz \\
7  & 16b &8MB &667MHz \\
\bottomrule
\end{tabular}
\end{table}




\subsubsection{Numerical Results}
We use a 2nd-order M-gPC with threshold 
$\varepsilon = 10^{-3}$ to build a surrogate model for the DRAM simulator. 
Due to the practical timing issue, we only run 200 MC samples to verify the effectiveness of the M-gPC model. 
\hzc{We solve \eqref{optimization_formula} to get 60 quadrature samples and weights for M-gPC in MATLAB with a 3.4GHz 8GB memory desktop, which takes 12.5 min. 
The time comparison between M-gPC modeling and MC for the first device configuration under 20\% uncertainty is shown in Table~\ref{tab:workloads}. Clearly, the cost of solving \eqref{optimization_formula} is negligible compared with the total simulation time.}  

\begin{figure}[htbp]
\centering
\subfigure{
\begin{minipage}[t]{0.5\linewidth}
\centering
\includegraphics[width=1.8in]{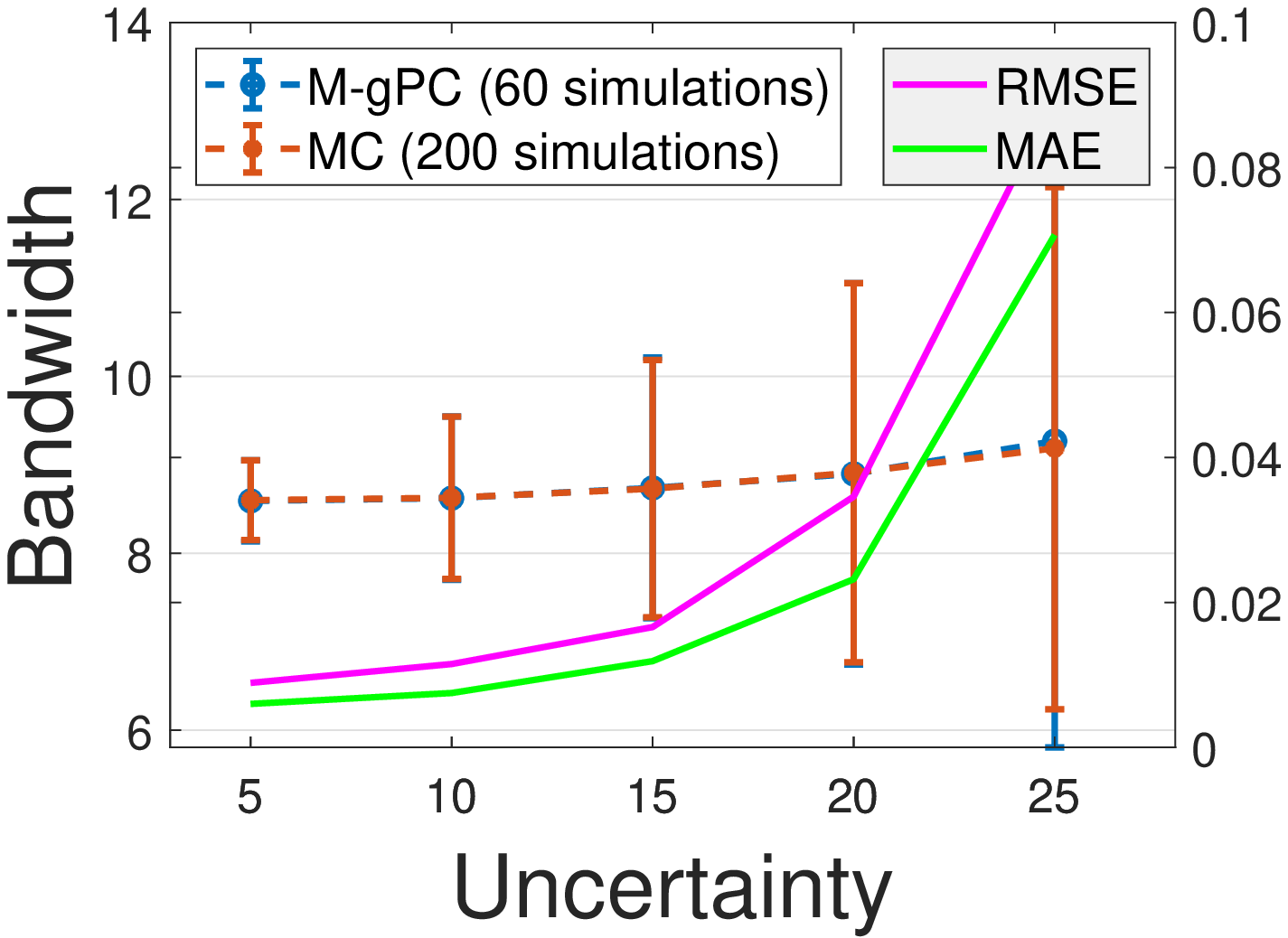}
\end{minipage}%
}%
\subfigure{
\begin{minipage}[t]{0.5\linewidth}
\centering
\includegraphics[width=1.8in]{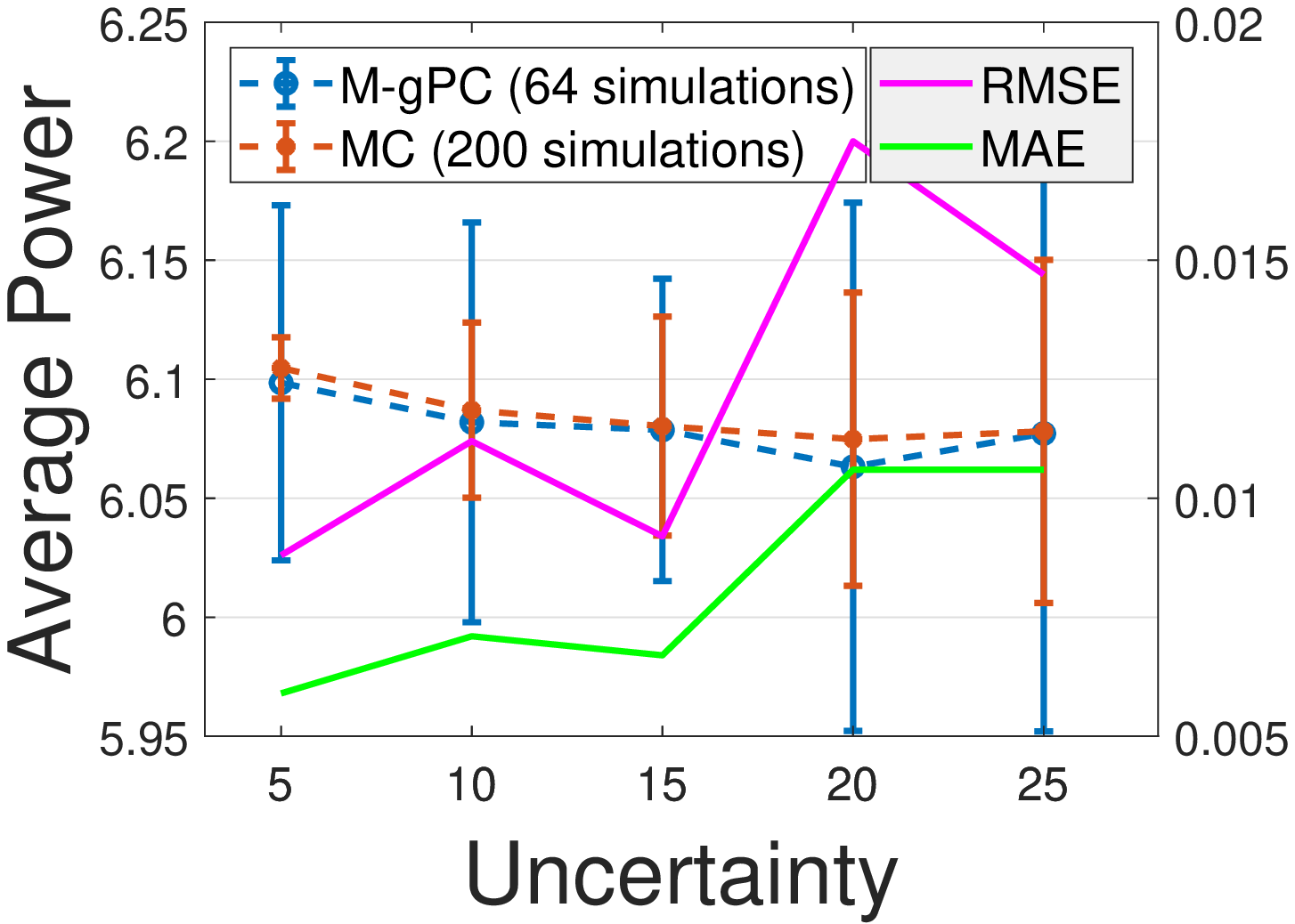}
\end{minipage}%
}%

\subfigure{
\begin{minipage}[t]{0.5\linewidth}
\centering
\includegraphics[width=1.8in]{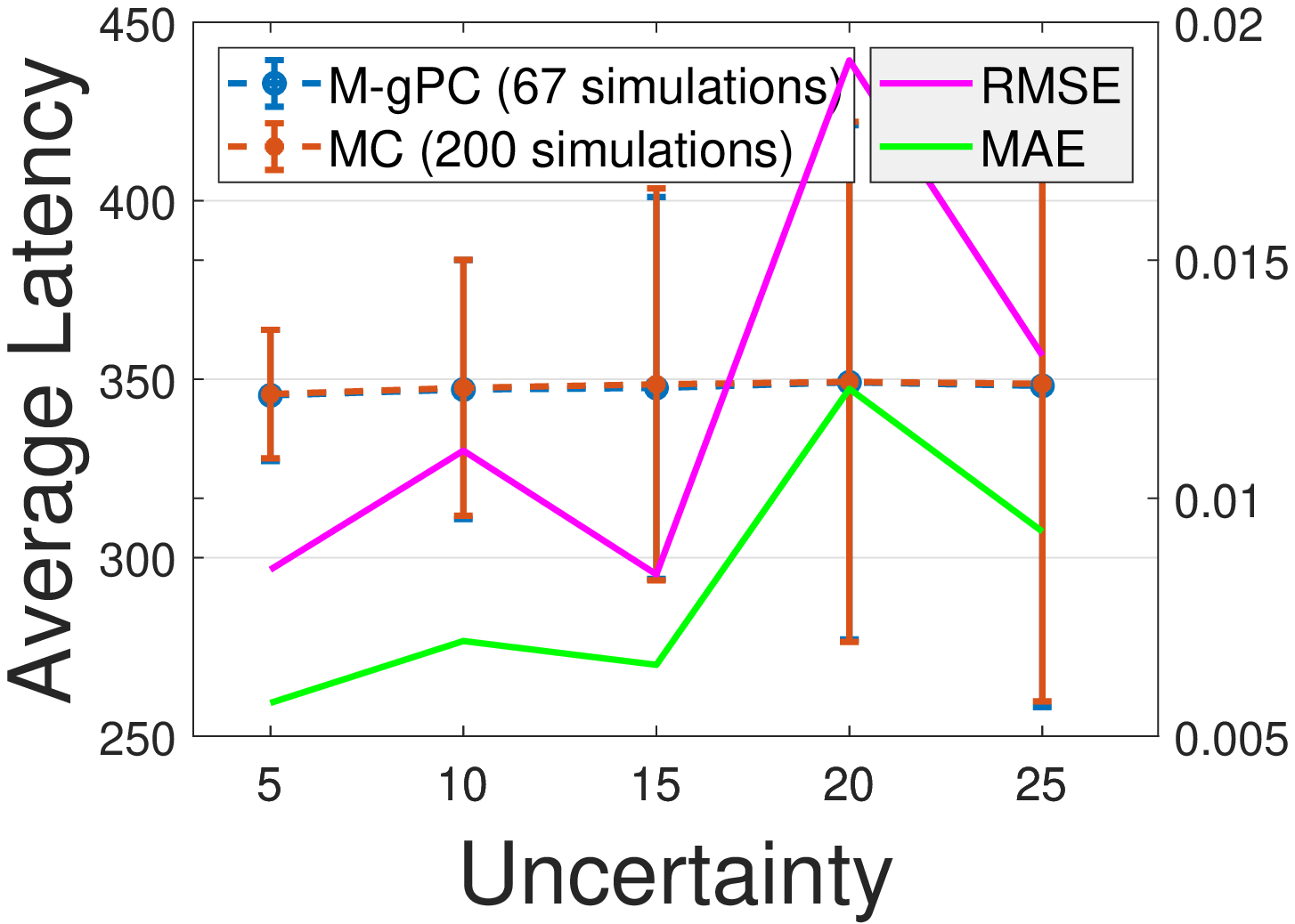}
\end{minipage}
}%
\subfigure{
\begin{minipage}[t]{0.5\linewidth}
\centering
\includegraphics[width=1.8in]{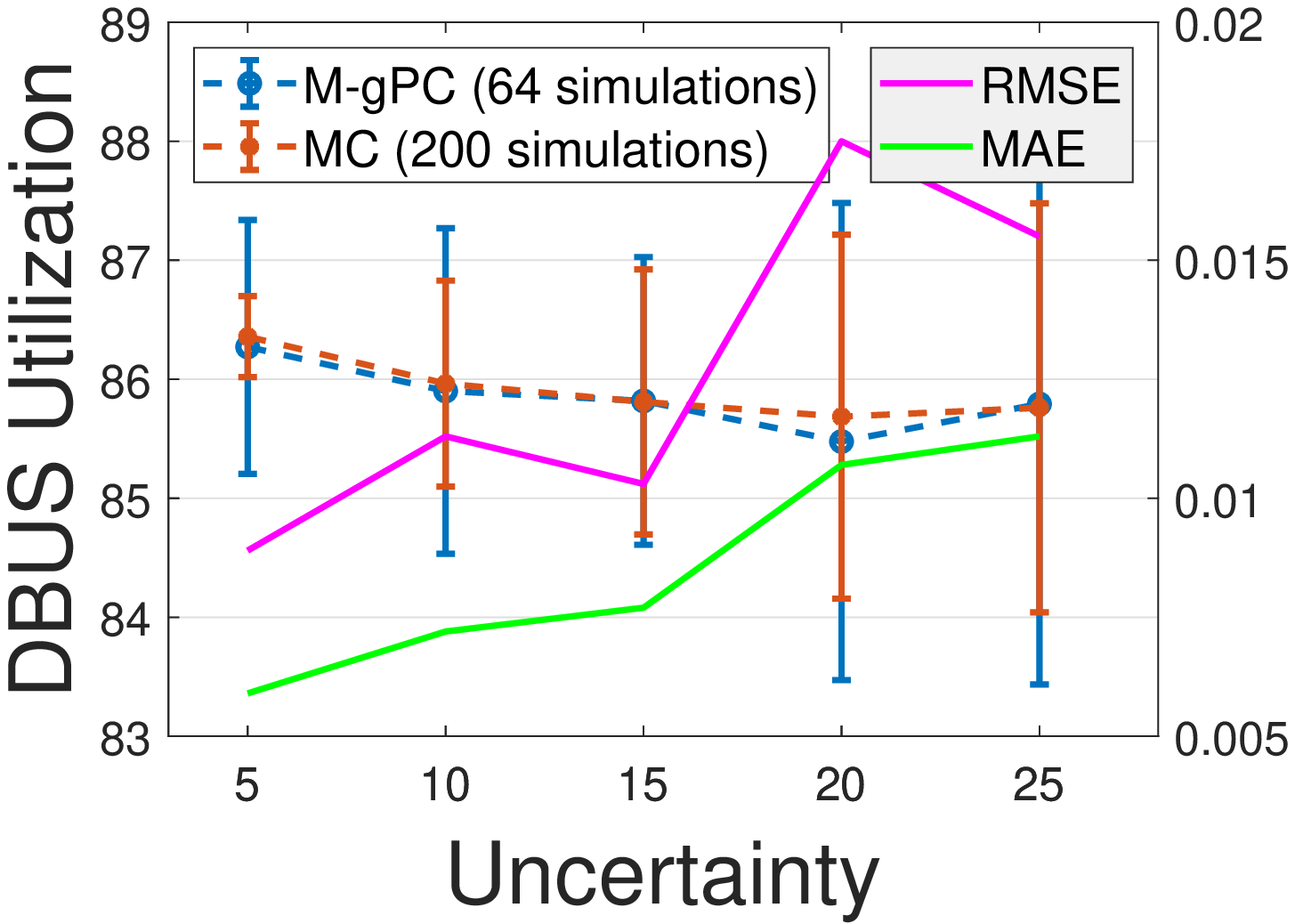}
\end{minipage}
}%
\centering
\caption{Performance under different uncertainty levels: M-gPC model (around 60 simulation samples) vs MC method (200 simulation samples).}\label{Fig:diff_uncertainty}
\end{figure}

\begin{figure*}[htbp]
\centering
\subfigure{
\begin{minipage}[t]{0.19\linewidth}
\centering
\includegraphics[width=1.4in]{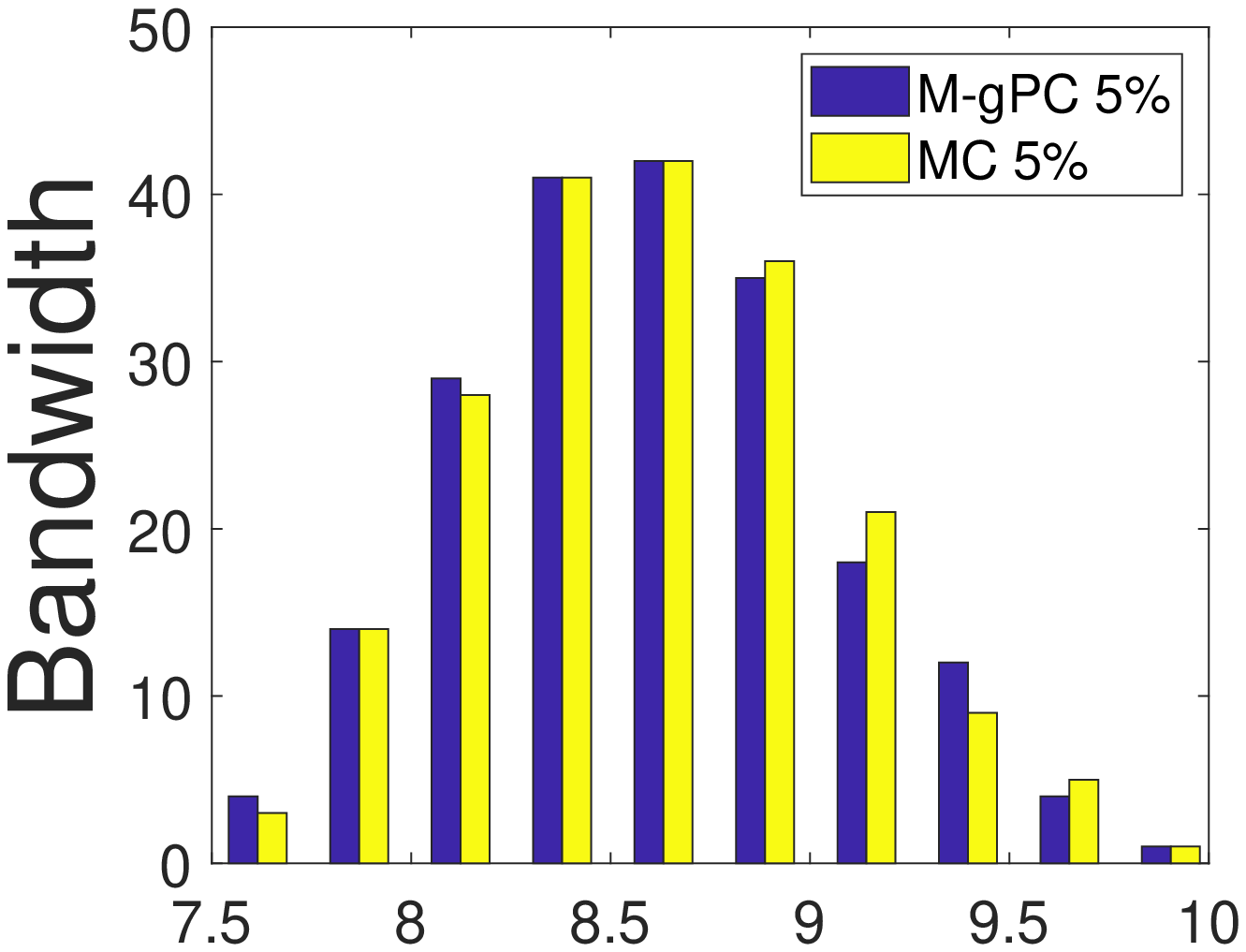}
\end{minipage}%
}%
\subfigure{
\begin{minipage}[t]{0.19\linewidth}
\centering
\includegraphics[width=1.4in]{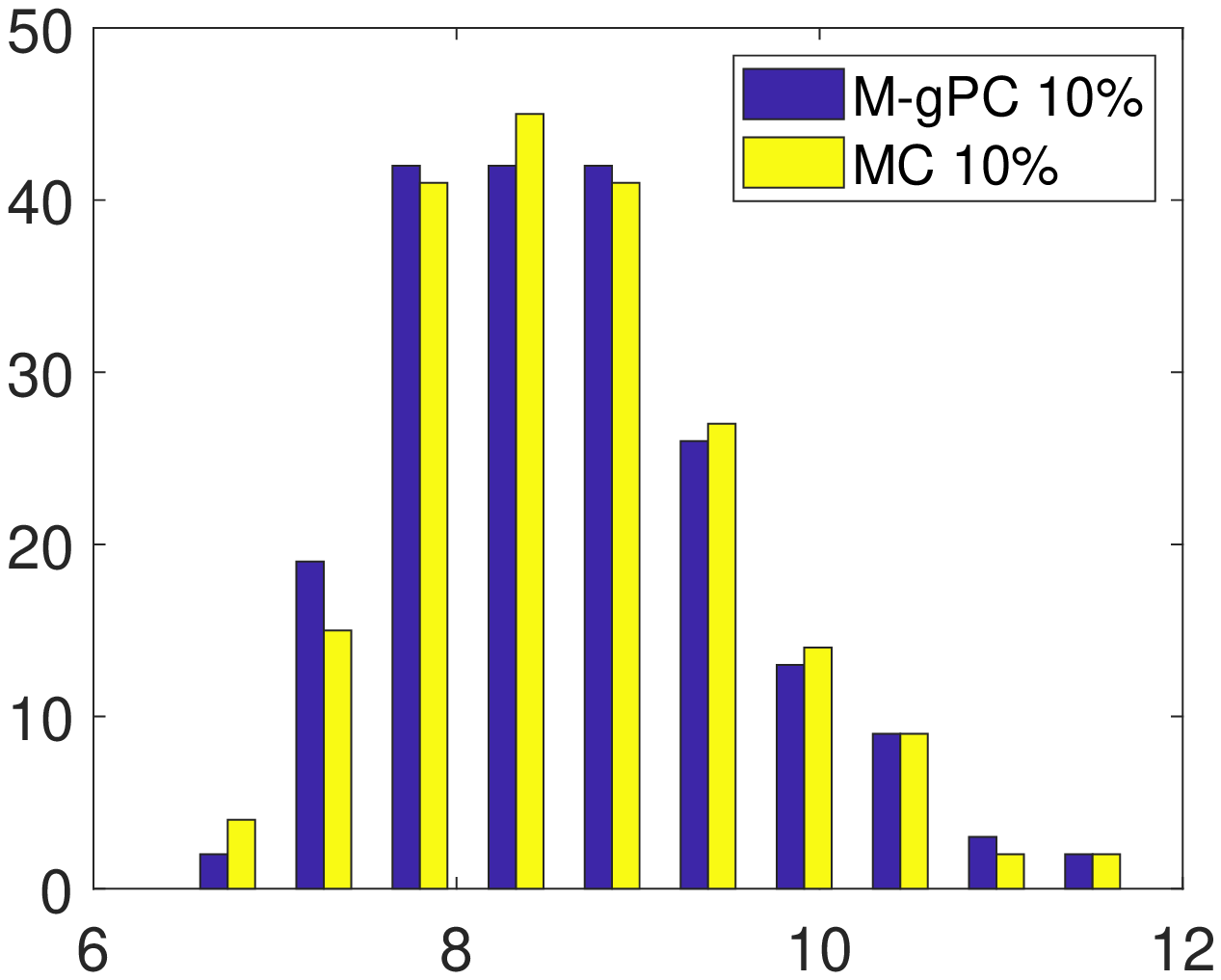}
\end{minipage}%
}%
\subfigure{
\begin{minipage}[t]{0.19\linewidth}
\centering
\includegraphics[width=1.4in]{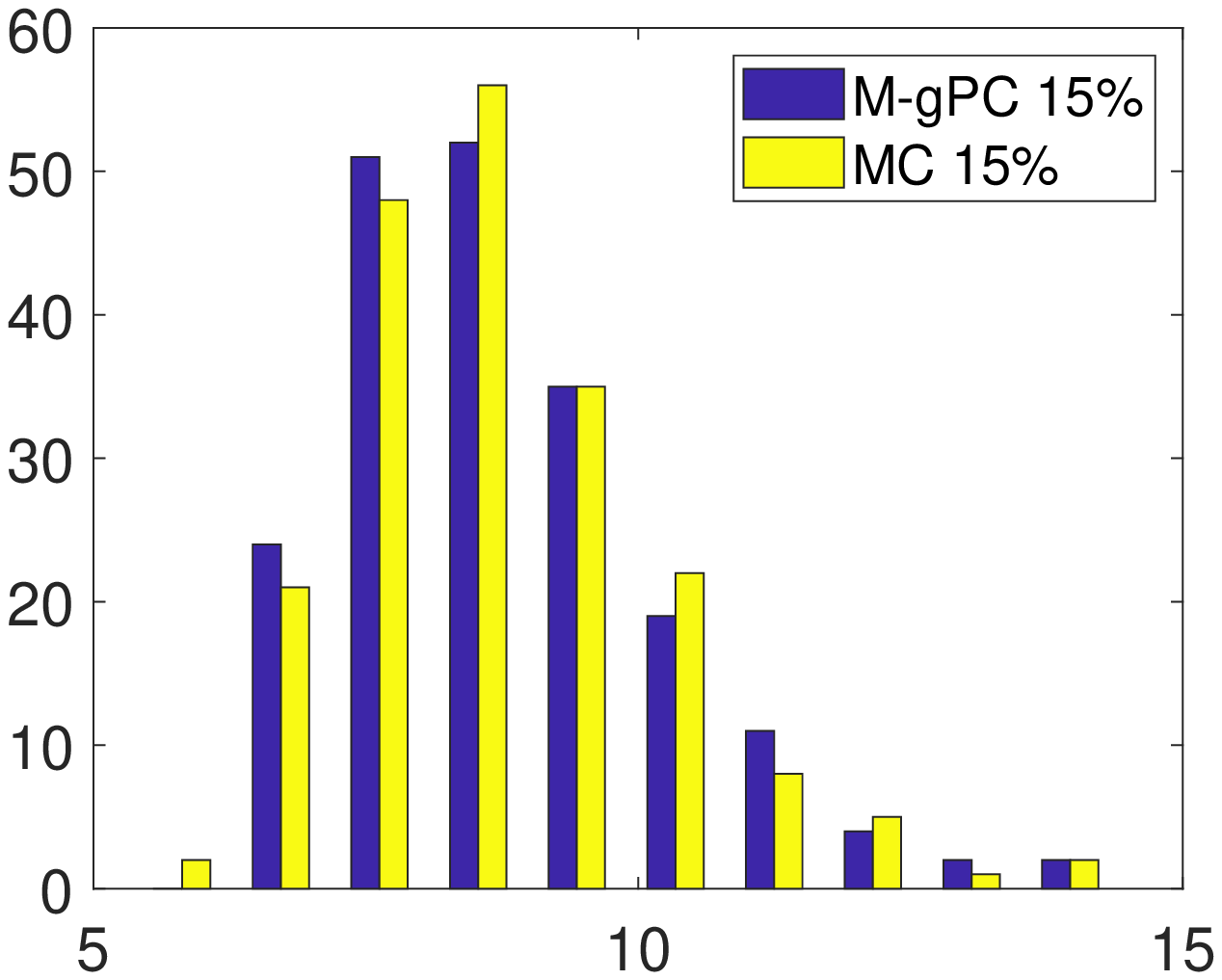}
\end{minipage}%
}%
\subfigure{
\begin{minipage}[t]{0.19\linewidth}
\centering
\includegraphics[width=1.4in]{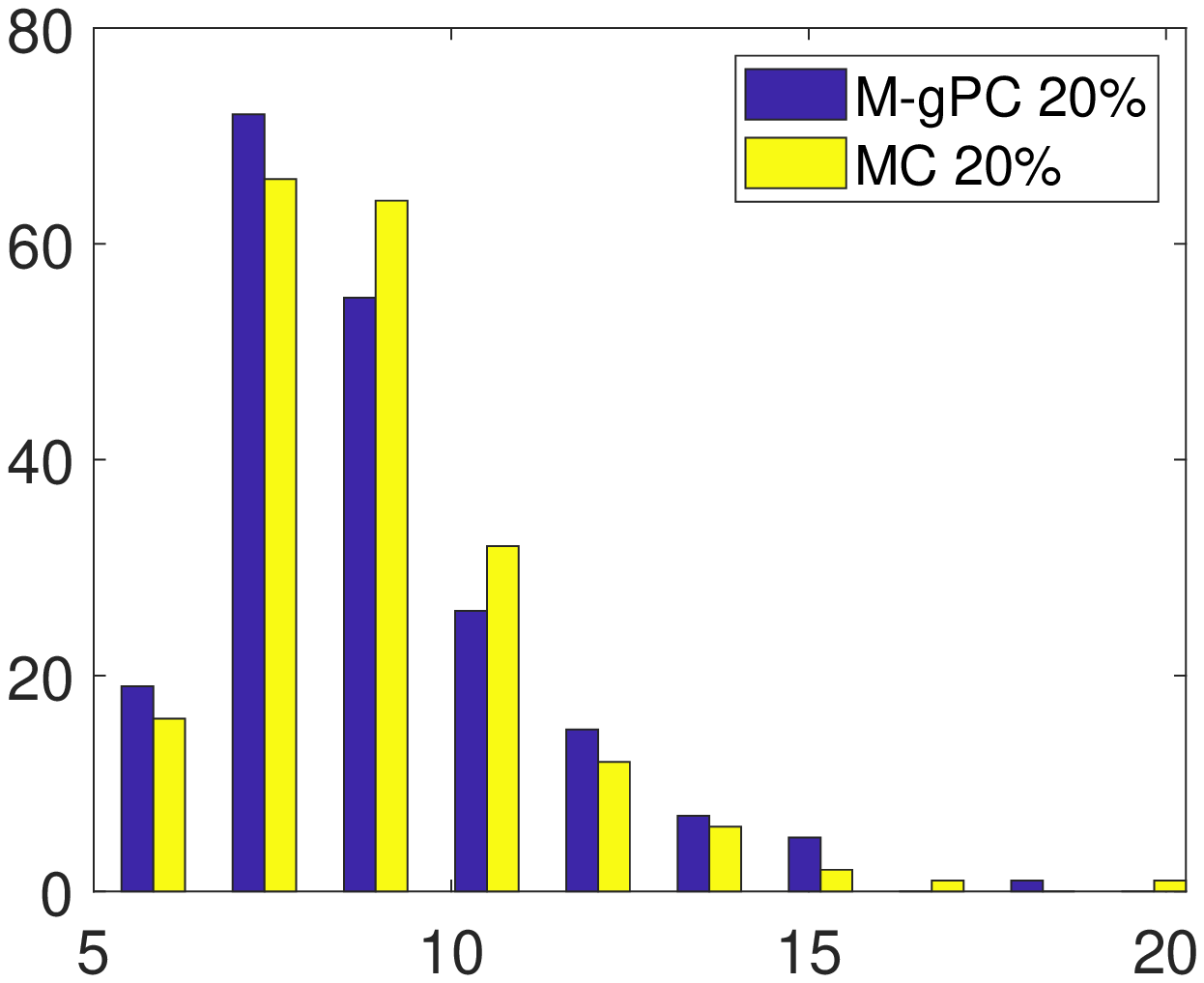}
\end{minipage}%
}%
\subfigure{
\begin{minipage}[t]{0.19\linewidth}
\centering
\includegraphics[width=1.4in]{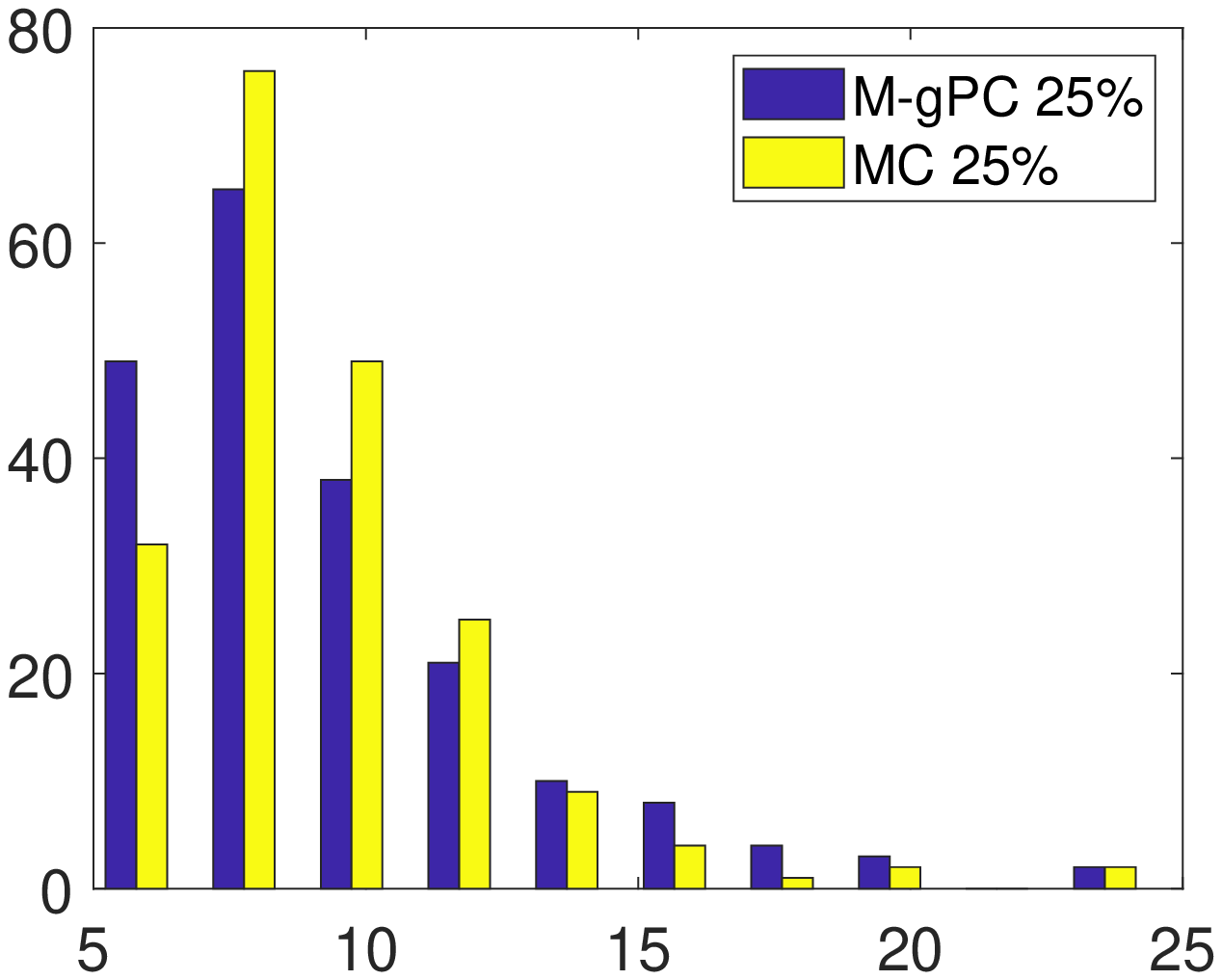}
\end{minipage}%
}%

\subfigure{
\begin{minipage}[t]{0.19\linewidth}
\centering
\includegraphics[width=1.4in]{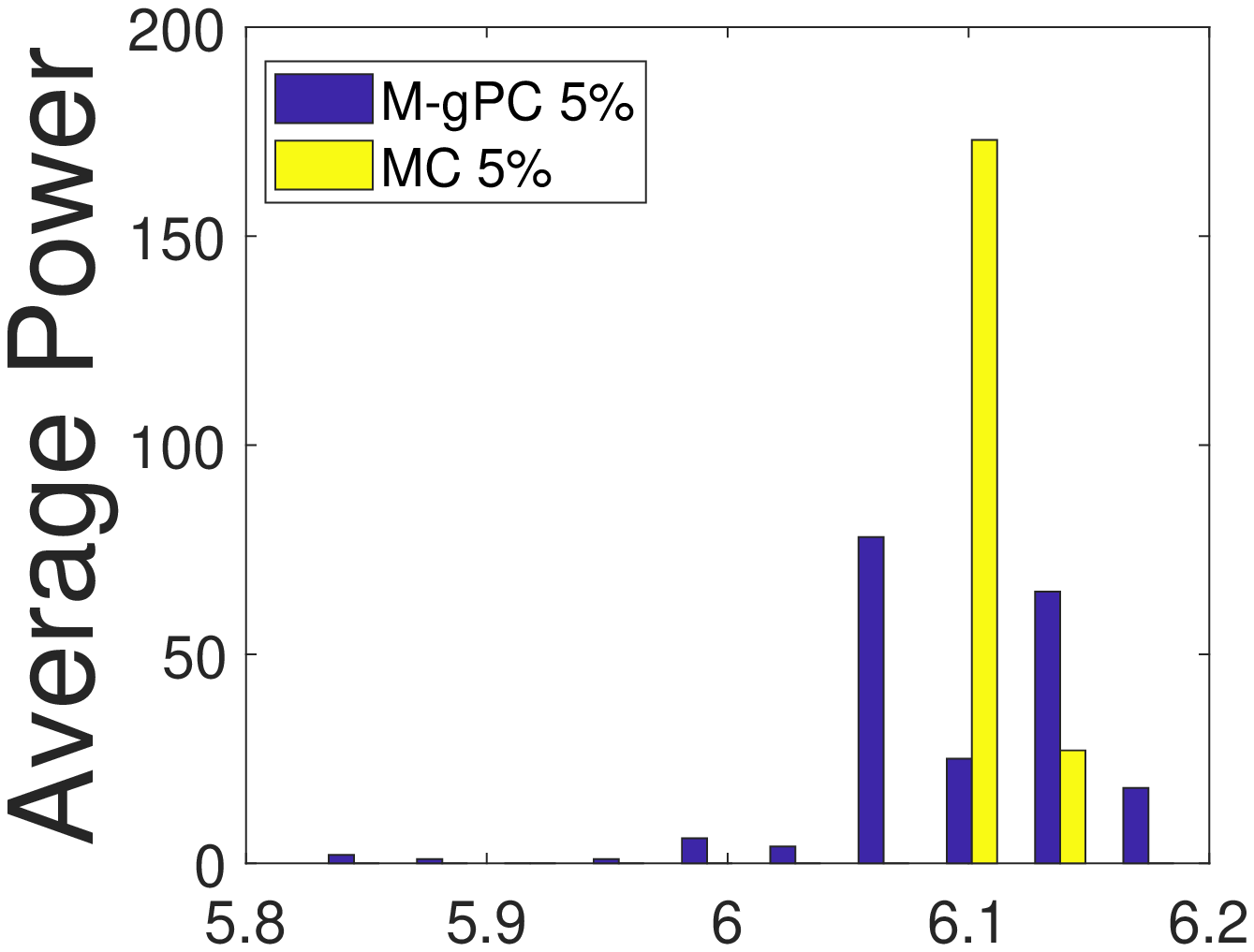}
\end{minipage}%
}%
\subfigure{
\begin{minipage}[t]{0.19\linewidth}
\centering
\includegraphics[width=1.4in]{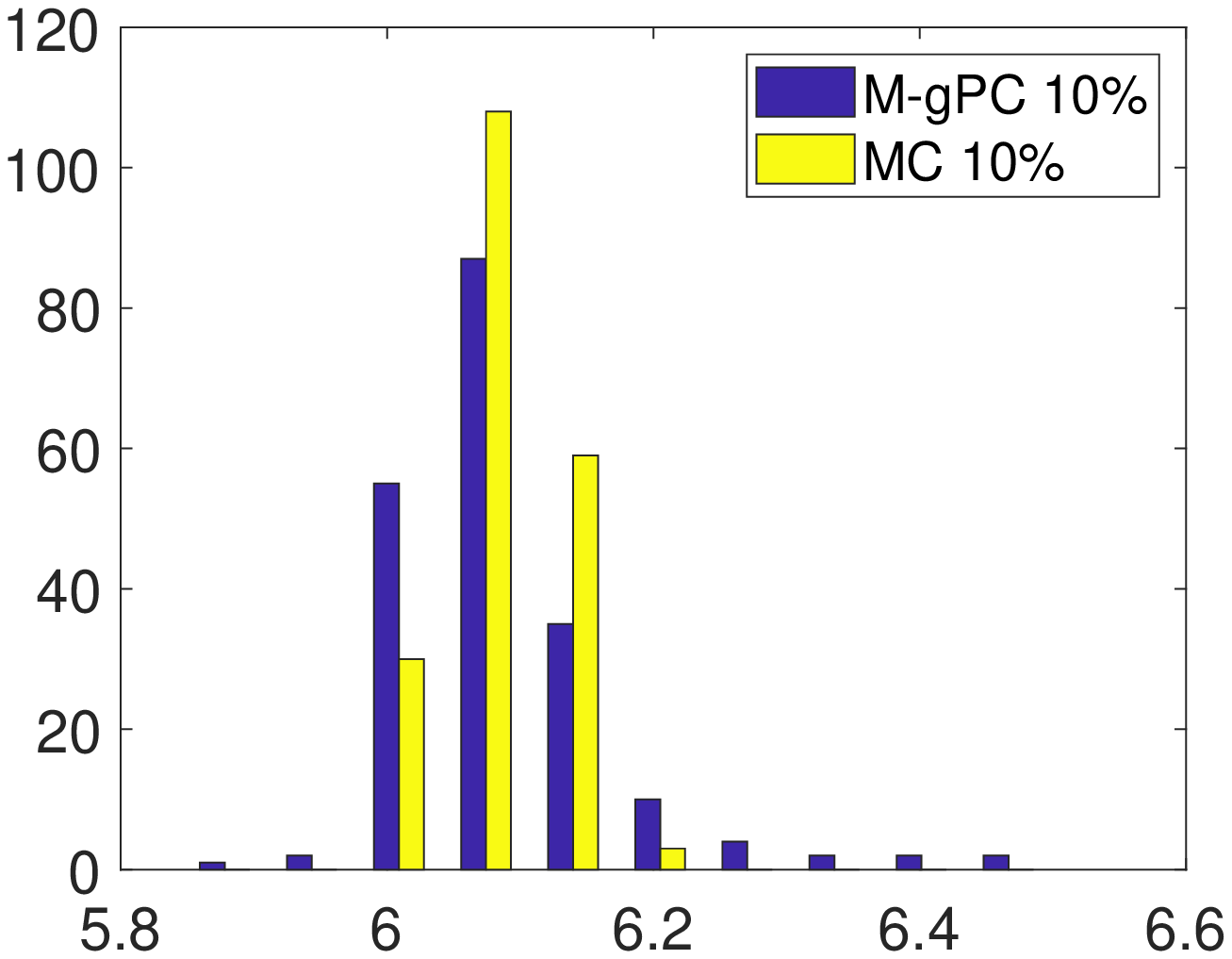}
\end{minipage}%
}%
\subfigure{
\begin{minipage}[t]{0.19\linewidth}
\centering
\includegraphics[width=1.4in]{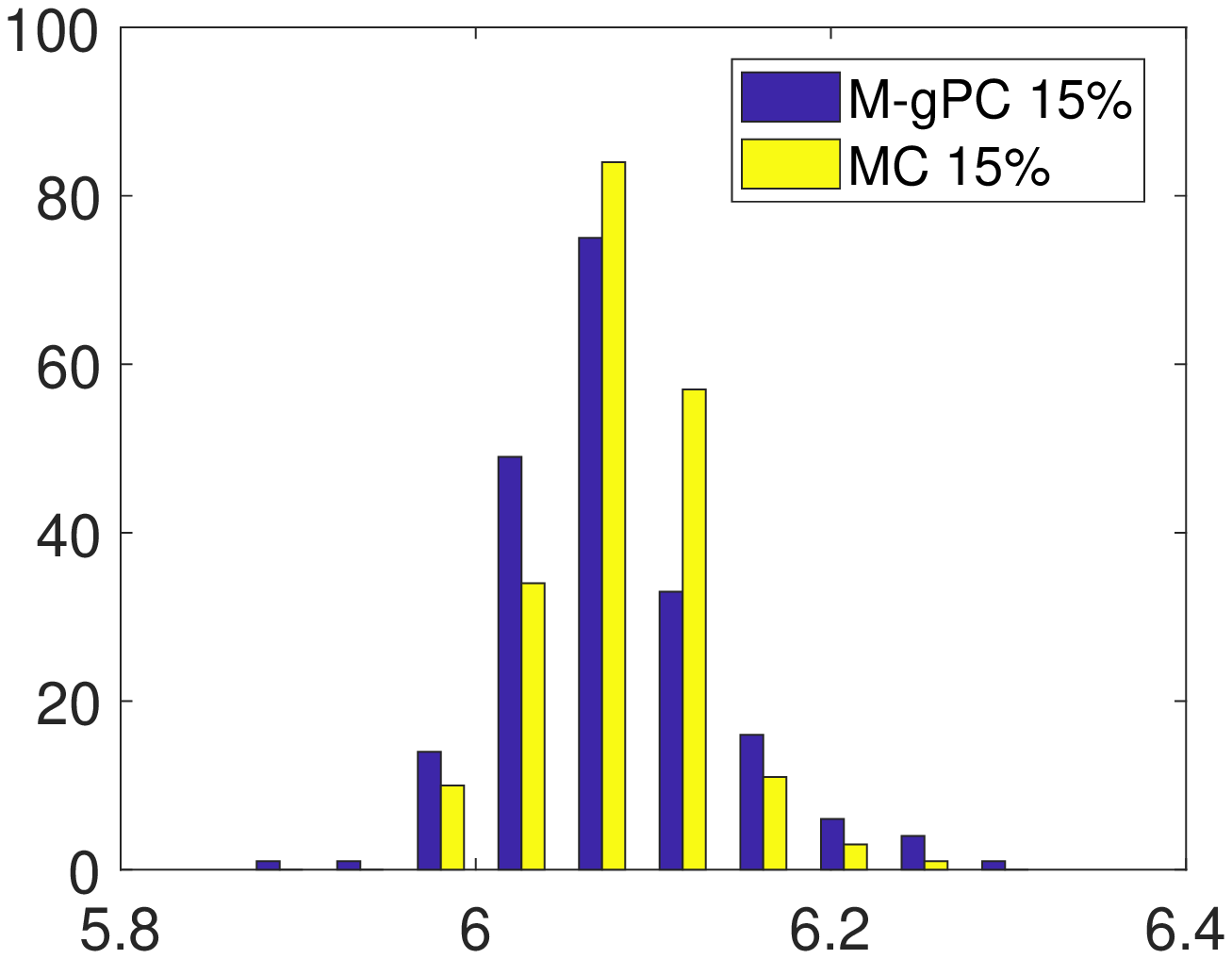}
\end{minipage}%
}%
\subfigure{
\begin{minipage}[t]{0.19\linewidth}
\centering
\includegraphics[width=1.4in]{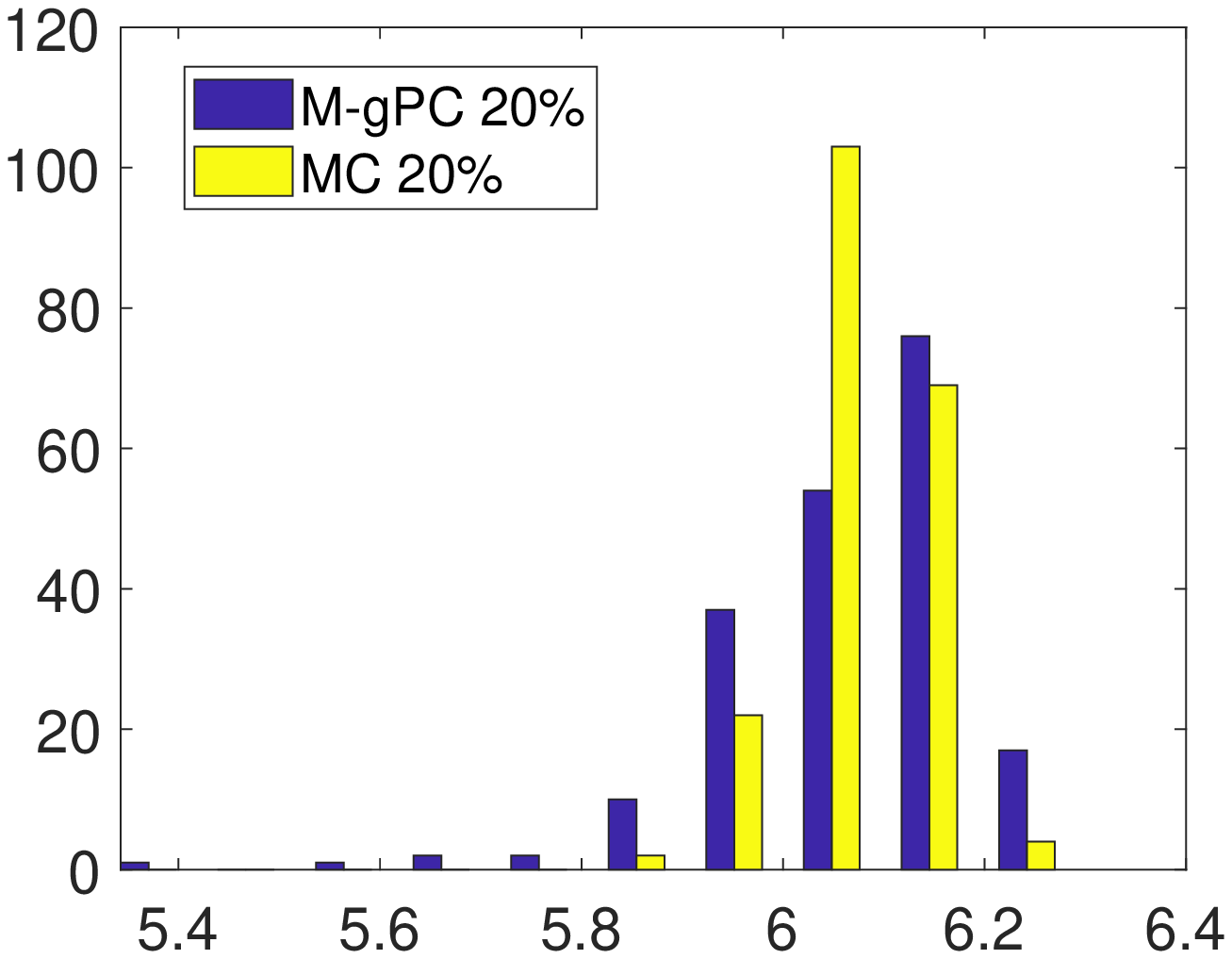}
\end{minipage}%
}%
\subfigure{
\begin{minipage}[t]{0.19\linewidth}
\centering
\includegraphics[width=1.4in]{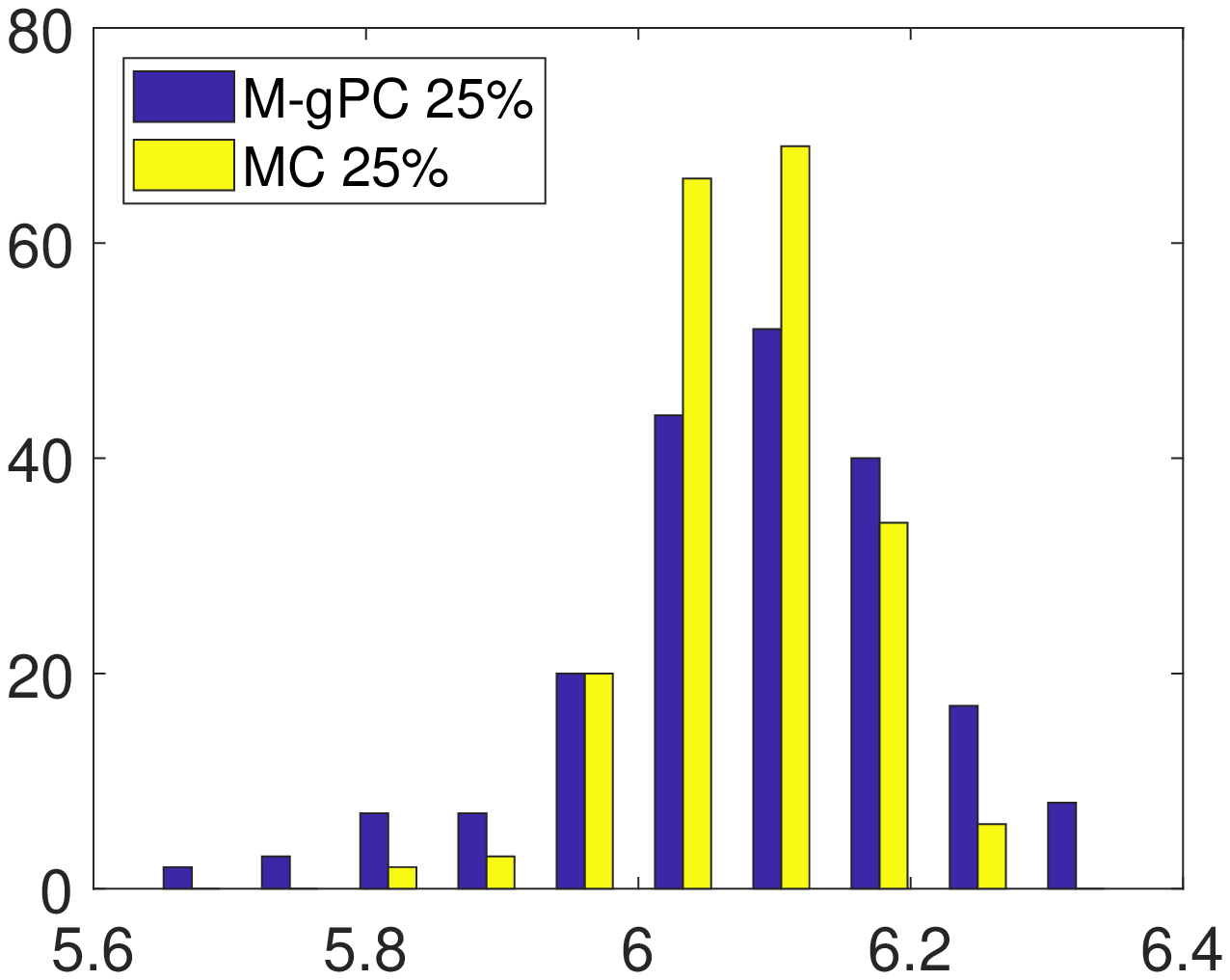}
\end{minipage}%
}%

\subfigure{
\begin{minipage}[t]{0.19\linewidth}
\centering
\includegraphics[width=1.4in]{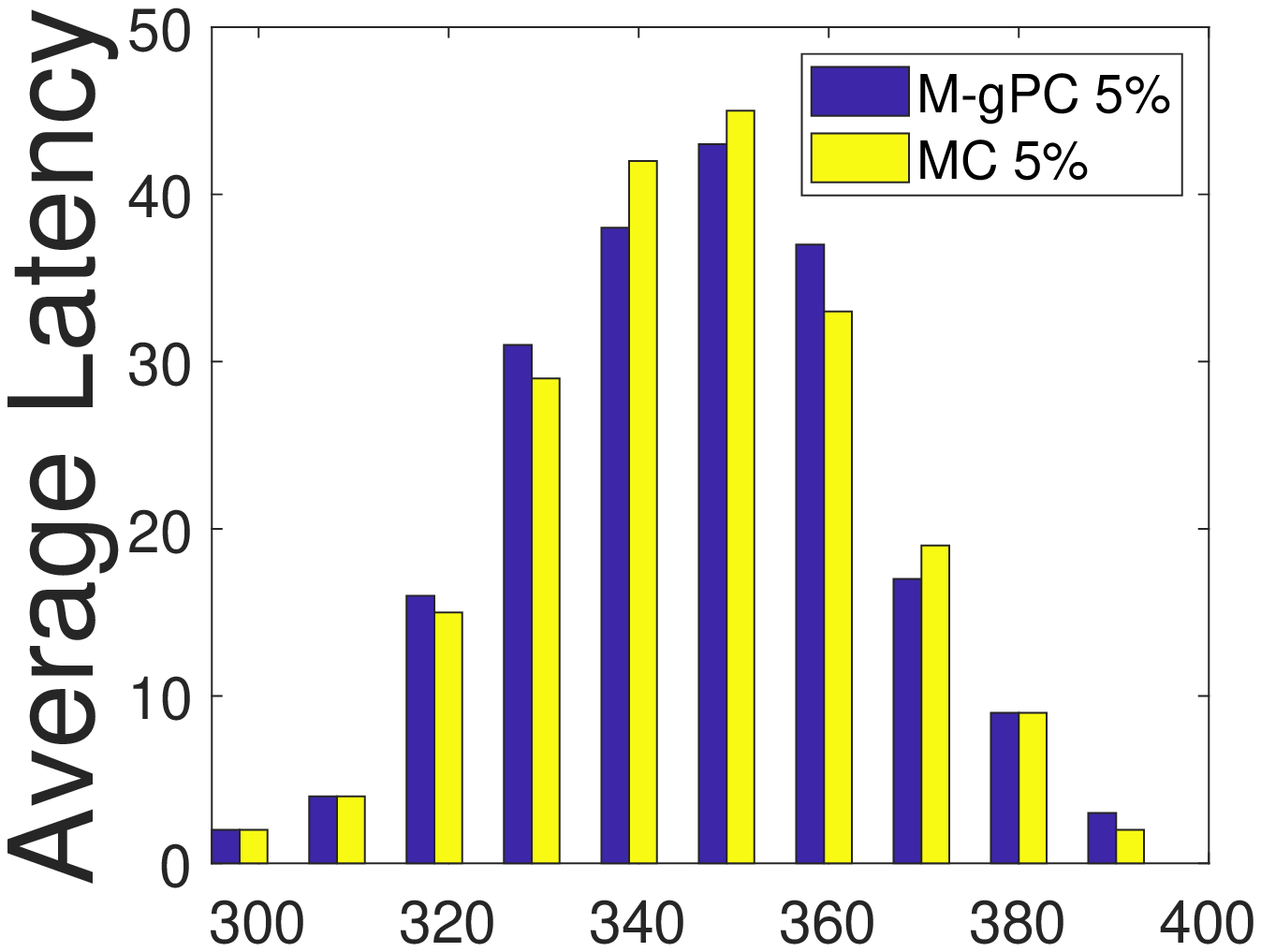}
\end{minipage}%
}%
\subfigure{
\begin{minipage}[t]{0.19\linewidth}
\centering
\includegraphics[width=1.4in]{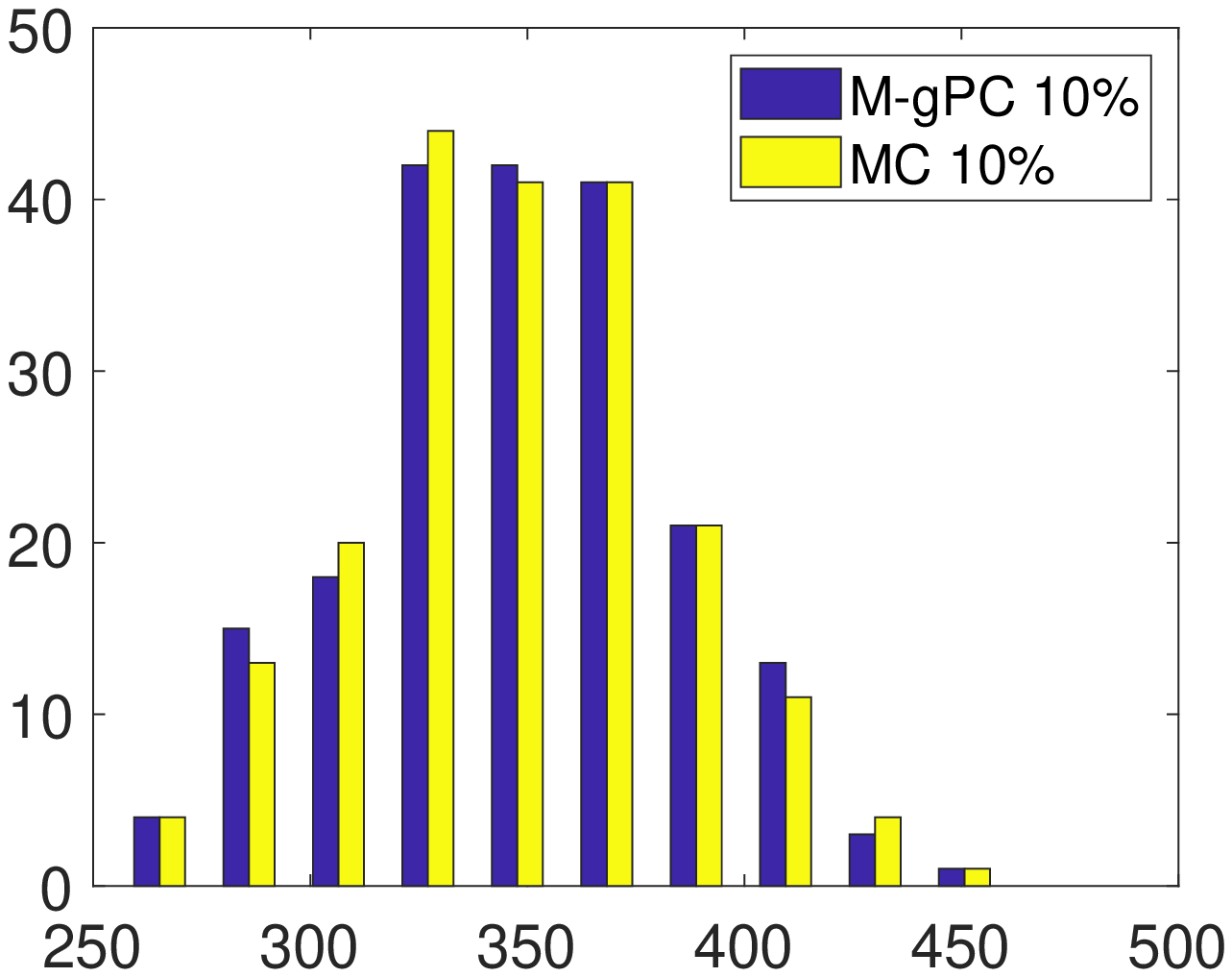}
\end{minipage}%
}%
\subfigure{
\begin{minipage}[t]{0.19\linewidth}
\centering
\includegraphics[width=1.4in]{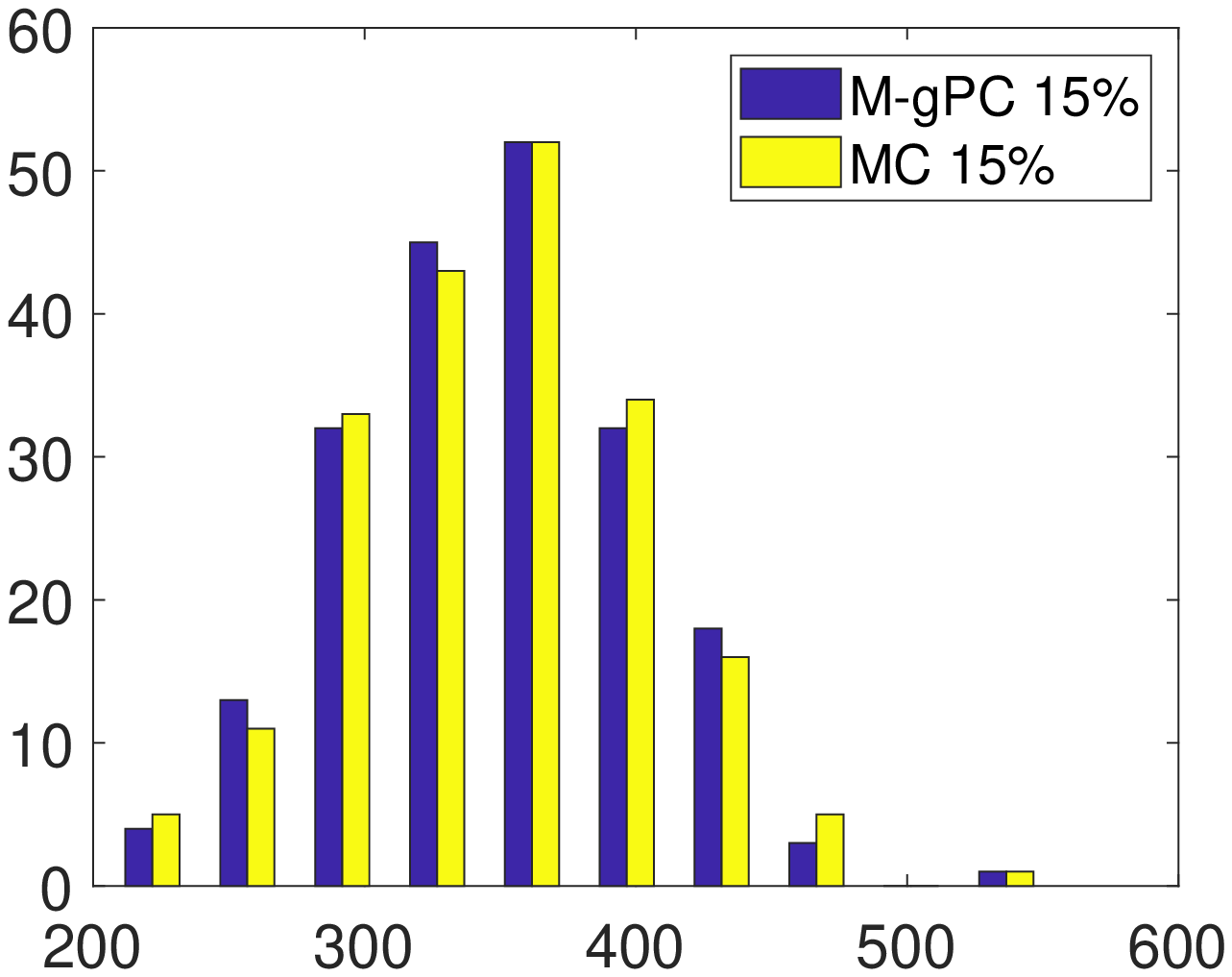}
\end{minipage}%
}%
\subfigure{
\begin{minipage}[t]{0.19\linewidth}
\centering
\includegraphics[width=1.4in]{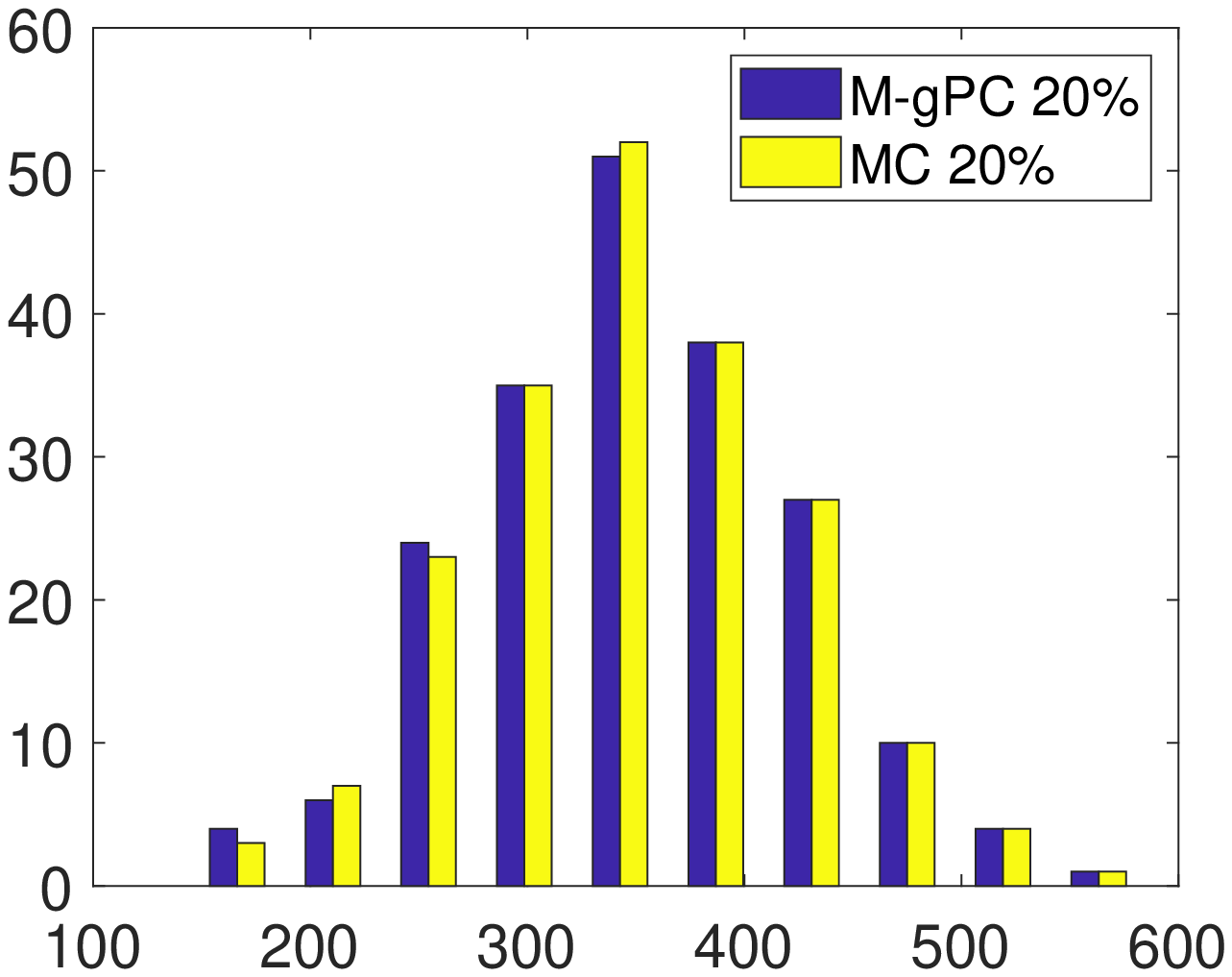}
\end{minipage}%
}%
\subfigure{
\begin{minipage}[t]{0.19\linewidth}
\centering
\includegraphics[width=1.4in]{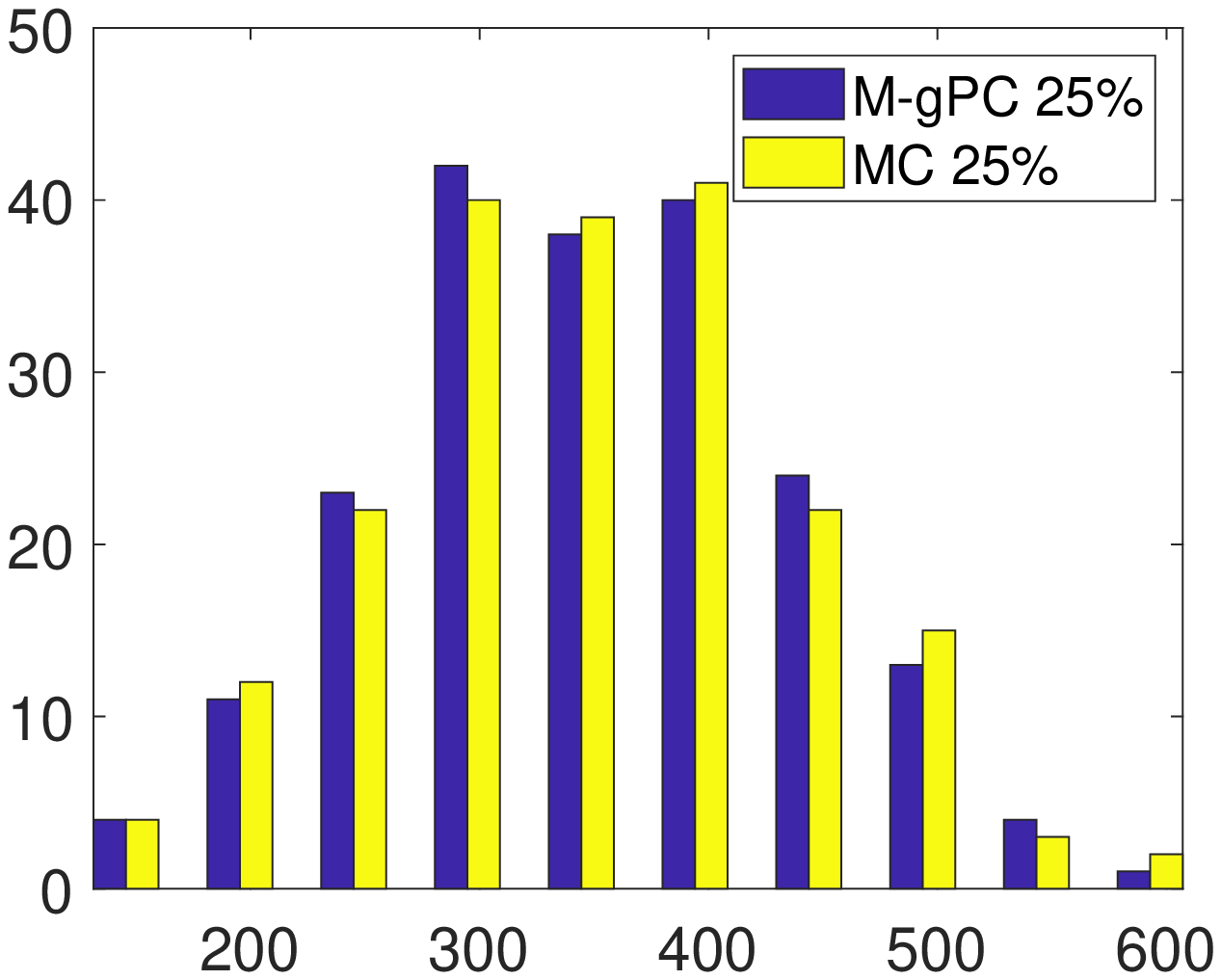}
\end{minipage}%
}%

\subfigure{
\begin{minipage}[t]{0.19\linewidth}
\centering
\includegraphics[width=1.4in]{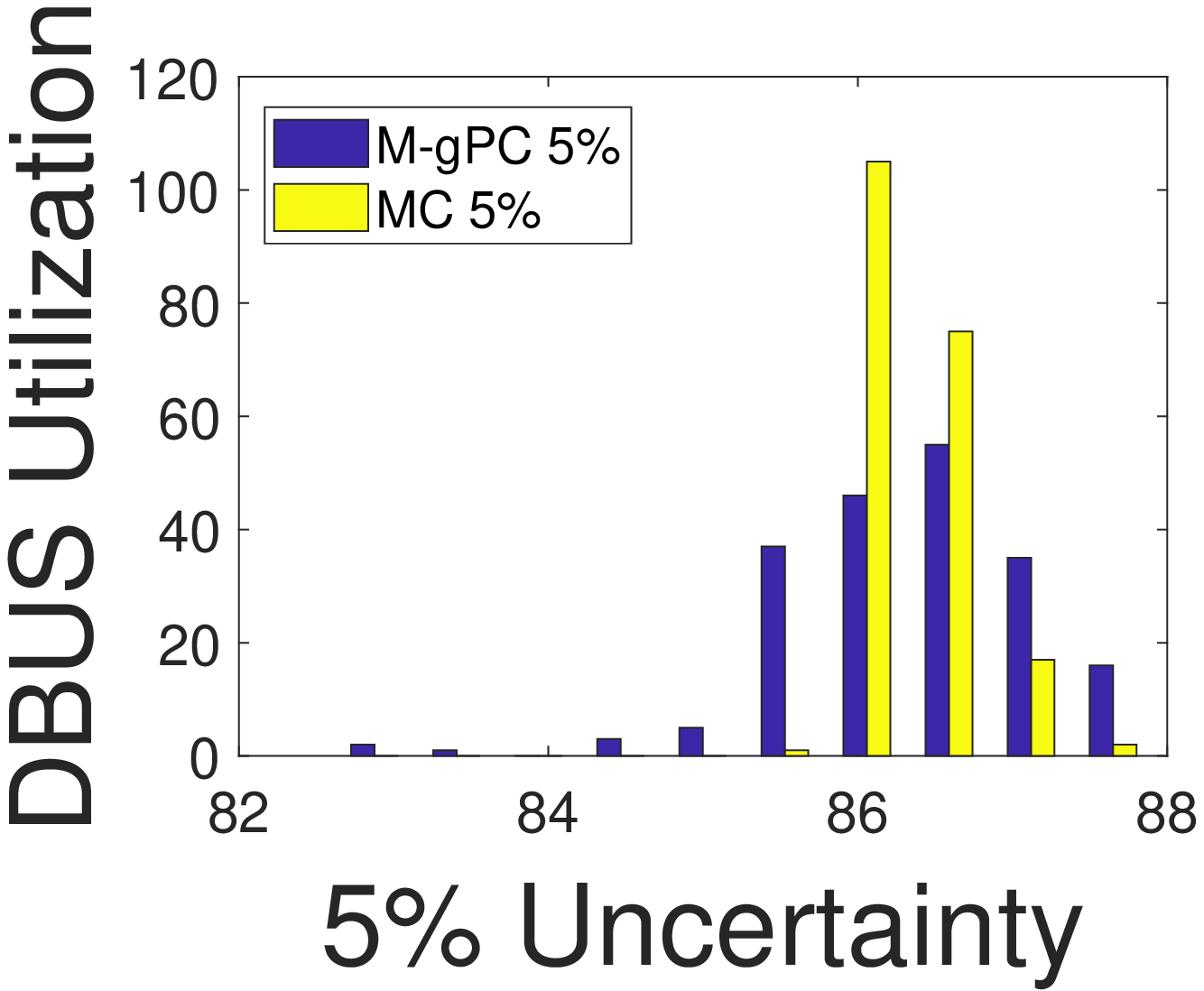}
\end{minipage}%
}%
\subfigure{
\begin{minipage}[t]{0.19\linewidth}
\centering
\includegraphics[width=1.4in]{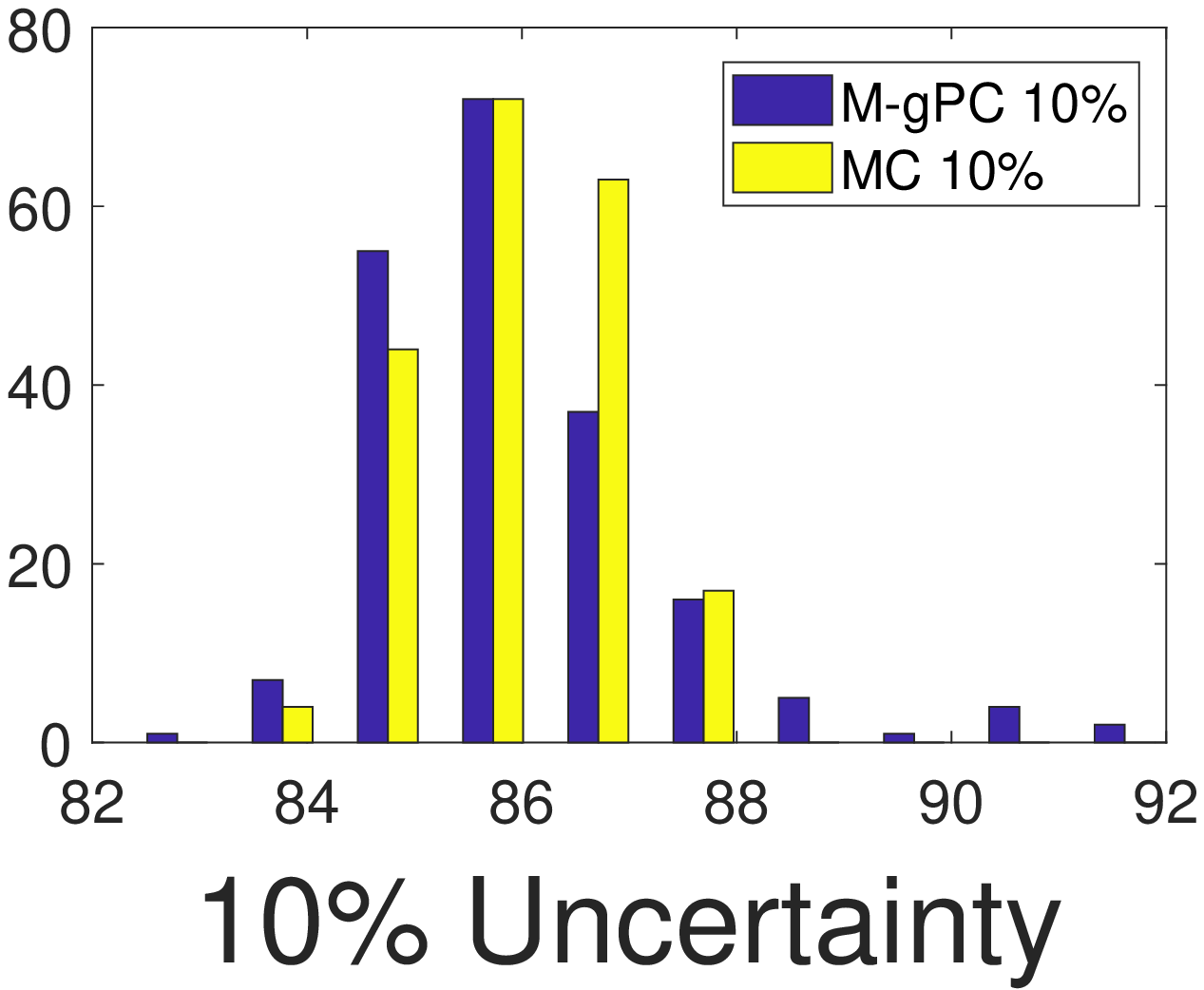}
\end{minipage}%
}%
\subfigure{
\begin{minipage}[t]{0.19\linewidth}
\centering
\includegraphics[width=1.4in]{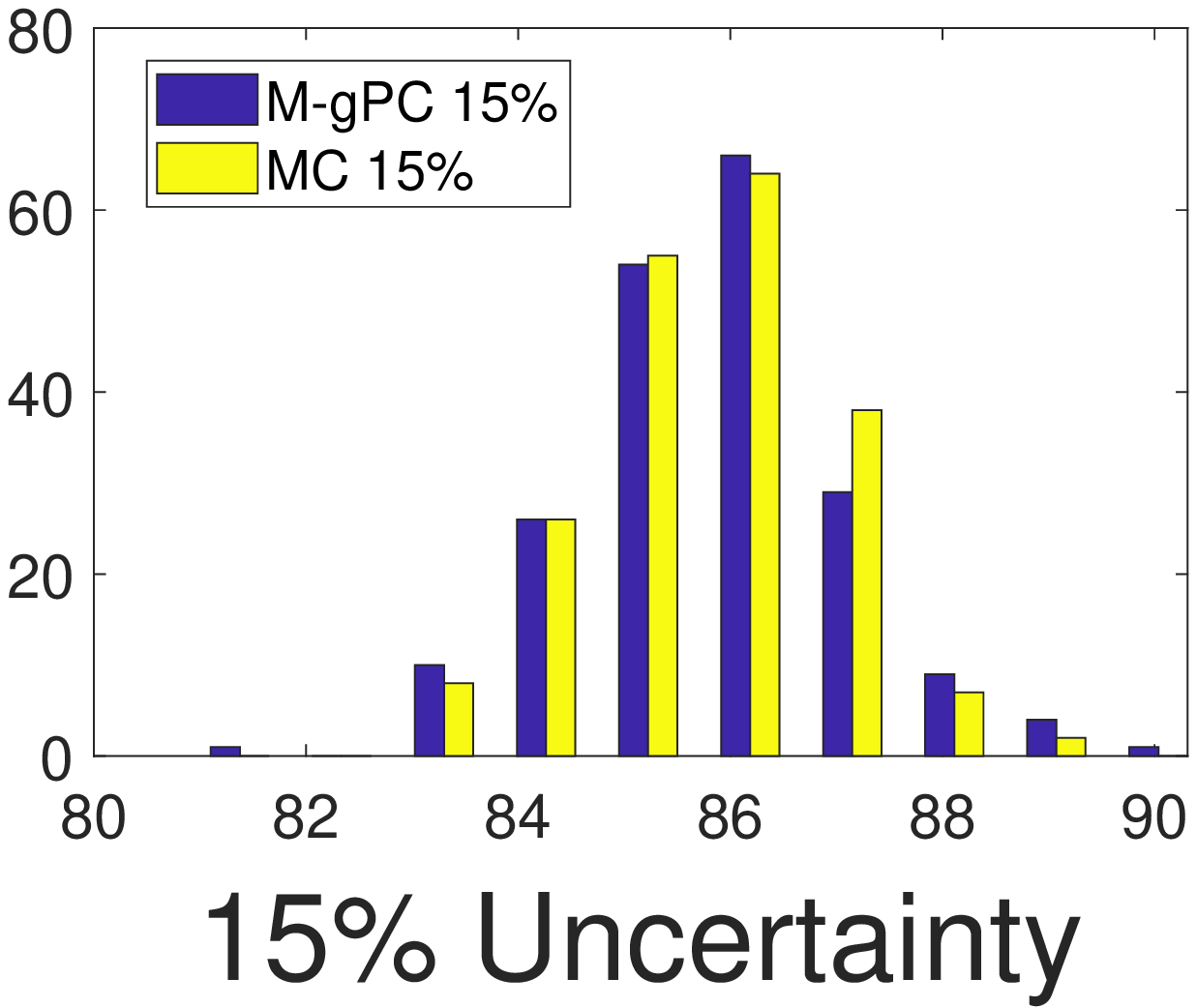}
\end{minipage}%
}%
\subfigure{
\begin{minipage}[t]{0.19\linewidth}
\centering
\includegraphics[width=1.4in]{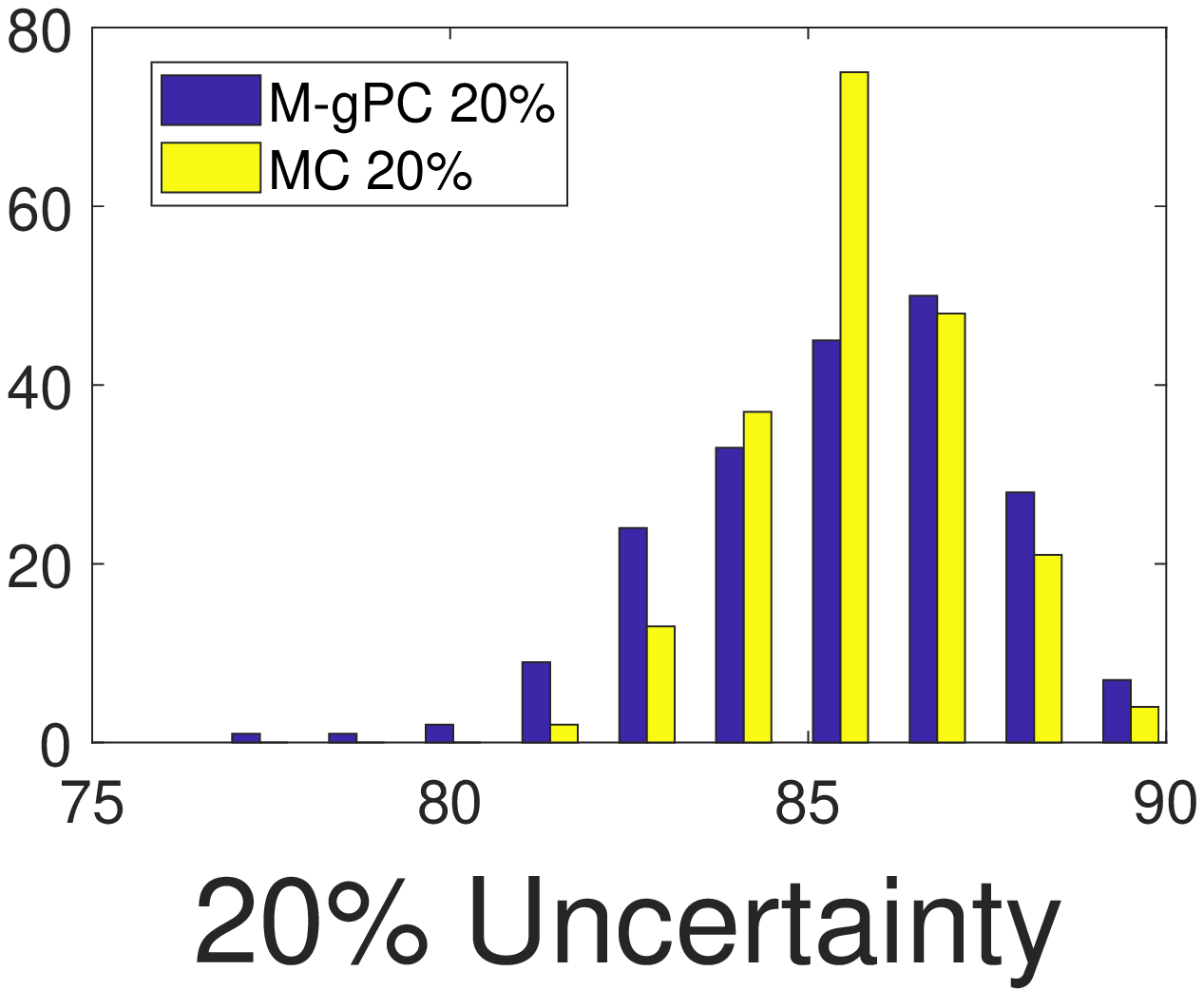}
\end{minipage}%
}%
\subfigure{
\begin{minipage}[t]{0.19\linewidth}
\centering
\includegraphics[width=1.4in]{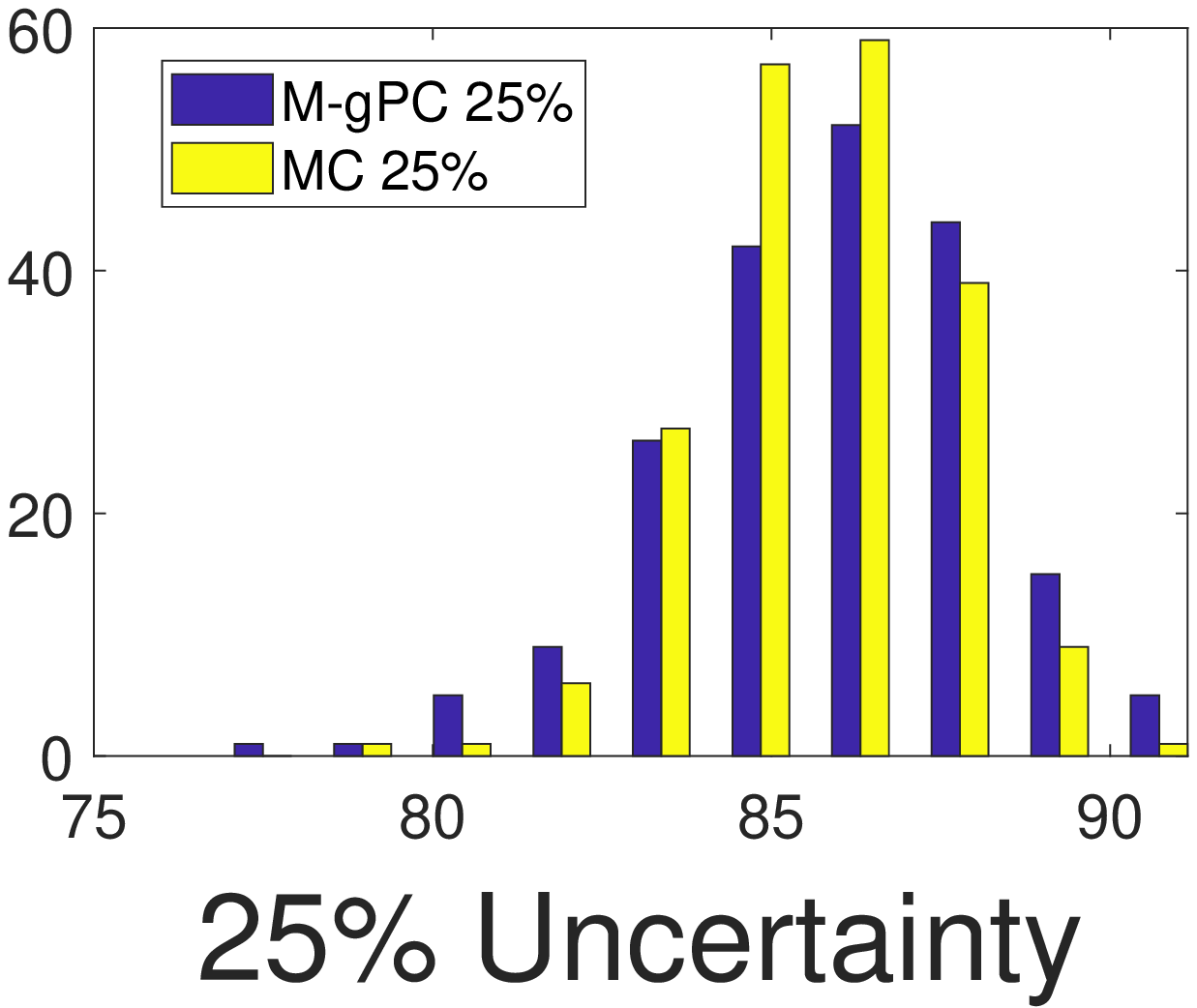}
\end{minipage}%
}%
\centering
\caption{Histograms of different performances under different uncertainty levels.}
\label{Fig:diff_uncertainty_hist}
\end{figure*}

\textbf{Performance under different uncertainty levels:}
in our setting, the mean value is set as the configuration without uncertainty, and the standard deviation $\sigma$ is set as
$\sigma  = \alpha  \cdot \mu$, where $\alpha$ is defined as the uncertainty level. 
For binomial distributions, we can also use Gaussian parameters to approximate it.
The performance of aggregate average bandwidth, average power, average latency and data bus (DBUS) utilization are illustrated in Fig. \ref{Fig:diff_uncertainty}. 
It is expected that the standard deviation will increase when the uncertainty level increases, which can be well captured by the M-gPC model with around only 60 samples. 
The sample number may vary since different input distributions lead to different M-gPC samples.  
The histograms of these four metrics are shown in Fig. \ref{Fig:diff_uncertainty_hist}. 
The M-gPC model can capture the performance distributions well. For the bandwidth, the approximating error may be more sensitive to uncertainty levels.  
This is because the bandwidth performance function appears to be an inverse curve, which is relatively harder for a 2nd-order M-gPC to approximate. In this case, we can increase the order 
to get better approximation.
These results verify that the M-gPC model is able to handle the cases under different levels of uncertainty.

\begin{figure}[htbp]
\centering
\subfigure{
\begin{minipage}[t]{0.47\linewidth}
\centering
\includegraphics[width=1.6in]{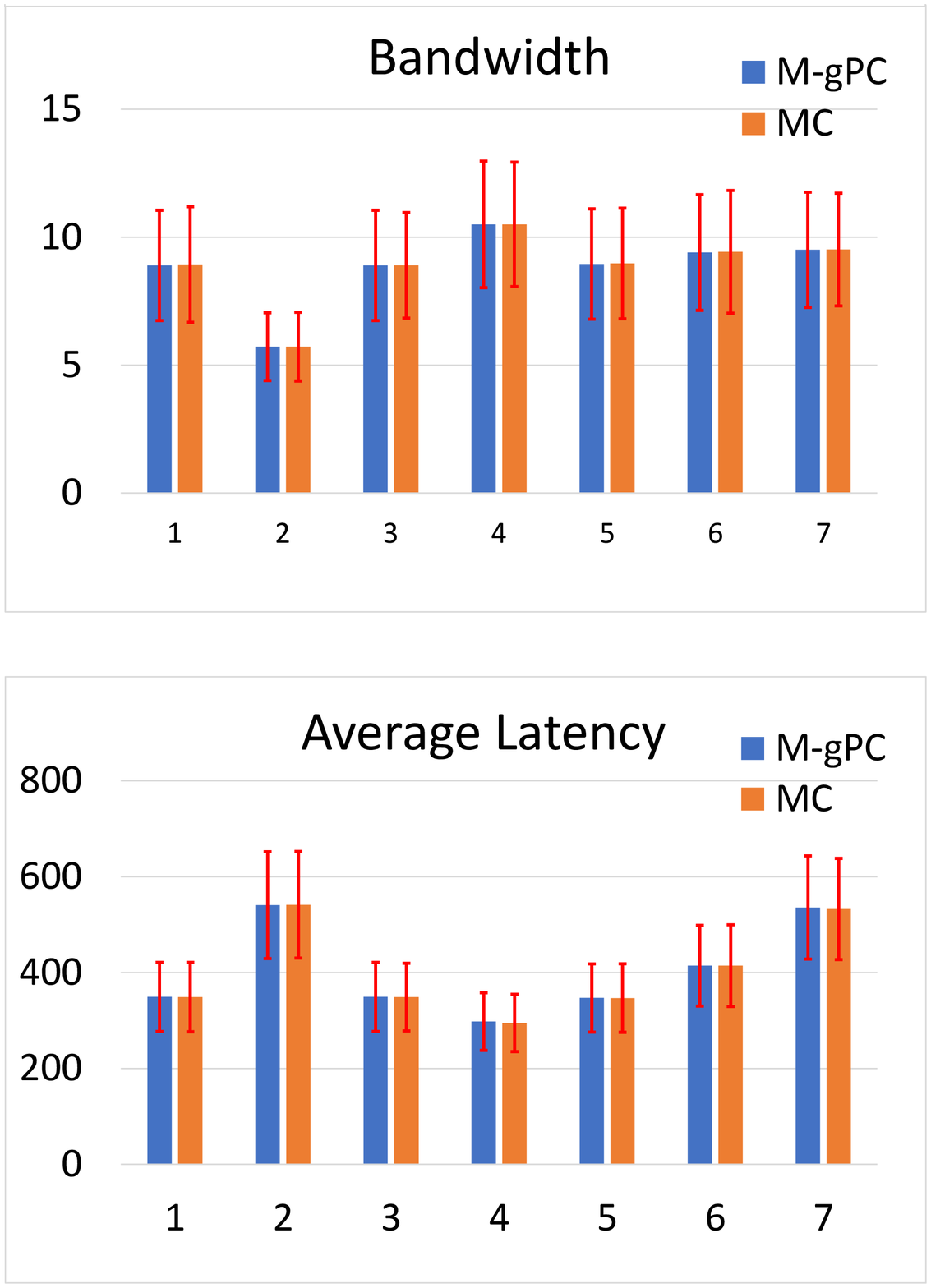}
\end{minipage}%
}%
\subfigure{
\begin{minipage}[t]{0.47\linewidth}
\centering
\includegraphics[width=1.6in]{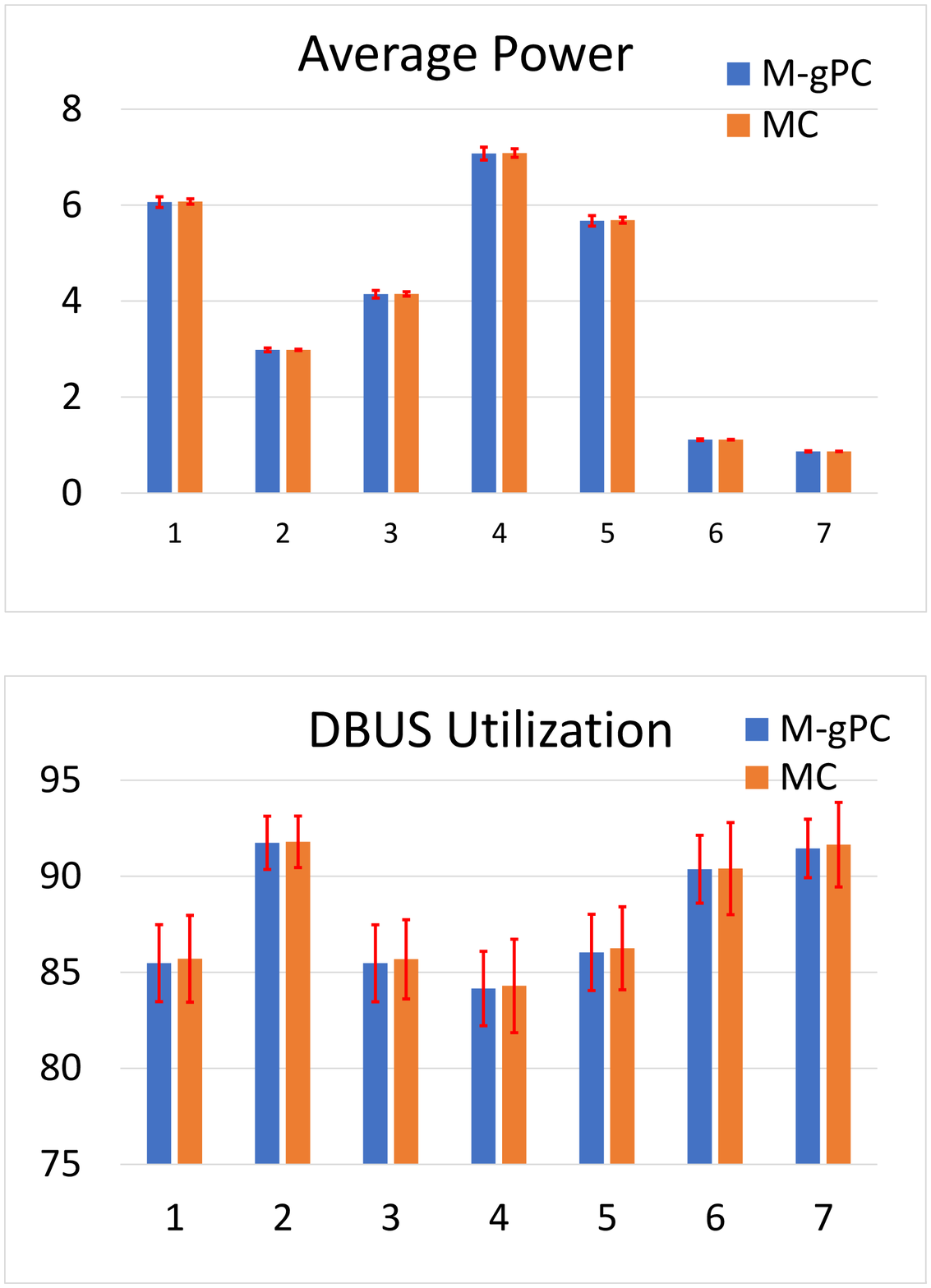}
\end{minipage}%
}%

\centering
\caption{Performance under different device configurations: M-gPC model (around 60 simulation samples) vs MC method (200 simulation samples).}\label{Fig_diff_config}
\end{figure}

\textbf{Performance under different device configurations:}
we run different device configurations with 20\% uncertainties on the 602.gcc trace. The results are illustrated in Fig. \ref{Fig_diff_config}, which show that, for different DRAM devices, M-gPC model can approximate the moments well with only around 60 samples. The errors for approximating the distribution are also small: RMSE varies in 1\%-4\%, MAE varies in 0.8\%-2.4\%.

\textbf{Performance under different workloads:}
we run the first device configuration with 20\% uncertainty on different workloads. 
The results are illustrated in Table \ref{Tb:diff_traces_results}. The moments and distribution are also approximated with a high accuracy, which shows the M-gPC model with only 66 simulation samples work well on different workloads too.
\begin{table*}[!h]
\caption{Performance under different workloads: M-gPC model (66 simulation samples) vs MC method (200 simulation samples).}
\label{Tb:diff_traces_results}
\centering
\small
\begin{tabular}{cccccccccc}
\toprule
& Workloads      & peribench      & gcc      & mcf      & xalancbmk      & x264      & deepsjeng      & leela      & specrand      \\
\hline
\multirow{6}{*}{\begin{tabular}[c]{@{}c@{}}Bandwidth\\ (GB/s)\end{tabular}}                                               & Mean (M-gPC) & 7.2847   & 8.9289   & 6.4874   & 7.8169   & 7.3144   & 7.1424   & 7.1413   & 7.2011   \\
                            & Mean (MC)  & 7.2834   & 8.9332   & 6.4868   & 7.8201   & 7.3146   & 7.1434   & 7.1422   & 7.2047   \\
                             & Std (M-gPC)  & 1.7545   & 2.0969   & 1.5718   & 1.8619   & 1.747    & 1.7194   & 1.7142   & 1.7318   \\
                                                                            & Std (MC)   & 1.7875   & 2.1479   & 1.6026   & 1.9002   & 1.7794   & 1.7538   & 1.7503   & 1.772    \\
                                                                            & RMSE       & 0.0318   & 0.0322   & 0.0319   & 0.0321   & 0.032    & 0.0319   & 0.0319   & 0.0319   \\
                            & MAE        & 0.0186   & 0.0185   & 0.0185   & 0.0185   & 0.0187   & 0.0185   & 0.0185   & 0.0187   \\
\hline
\multirow{6}{*}{\begin{tabular}[c]{@{}c@{}}Average\\ Power (watts)\end{tabular}}    & Mean (M-gPC) & 5.6782   & 6.0804   & 5.3127   & 5.7486   & 5.3717   & 5.5257   & 5.4308   & 5.5089   \\
                                                                            & Mean (MC)  & 5.6796   & 6.0831   & 5.3149   & 5.7517   & 5.3731   & 5.5281   & 5.4333   & 5.5112   \\
                                                                            & Std (M-gPC)  & 0.2159   & 0.1362   & 0.2168   & 0.1832   & 0.1871   & 0.2073   & 0.1934   & 0.2113   \\
                                                                            & Std (MC)   & 0.18     & 0.0817   & 0.1858   & 0.1428   & 0.1486   & 0.1732   & 0.1605   & 0.1773   \\
                                                                            & RMSE       & 0.0123   & 0.0124   & 0.0121   & 0.0122   & 0.0123   & 0.0121   & 0.0121   & 0.0125   \\
                                                                        & MAE        & 0.0094   & 0.0096   & 0.0093   & 0.0093   & 0.0094   & 0.0093   & 0.0093   & 0.0096   \\
\hline
\multirow{6}{*}{\begin{tabular}[c]{@{}c@{}}Average \\ Latency (ns)\end{tabular}} & Mean (M-gPC) & 423.5401 & 347.5645 & 462.5911 & 393.1547 & 431.4108 & 438.715  & 424.1744 & 428.8241 \\
                                                                            & Mean (MC)  & 424.0171 & 348.0141 & 463.0697 & 393.5325 & 432.0178 & 439.1198 & 424.5305 & 429.1871 \\
                                                                            & Std (M-gPC)  & 90.7912  & 71.9639  & 99.6712  & 82.6582  & 91.3793  & 93.9994  & 90.5722  & 91.6698  \\
                                                                            & Std (MC)   & 89.3849  & 71.4477  & 97.9099  & 81.544   & 90.3919  & 92.4259  & 89.0638  & 90.7059  \\
                                                                            & RMSE       & 0.0129   & 0.013    & 0.0129   & 0.013    & 0.0131   & 0.0129   & 0.013    & 0.0132   \\
                                                                            & MAE        & 0.0101   & 0.0101   & 0.01     & 0.0103   & 0.0102   & 0.0101   & 0.0102   & 0.0103   \\
\hline
\multirow{6}{*}{\begin{tabular}[c]{@{}c@{}}DBUS\\ Utilization (\%)\end{tabular}} & Mean (M-gPC) & 70.0493  & 85.8639  & 62.3914  & 75.1677  & 70.3477  & 68.6842  & 68.6735  & 69.2553  \\
& Mean (MC)  & 70.0251  & 85.8725  & 62.3752  & 75.1851  & 70.333   & 68.6808  & 68.669   & 69.2661  \\
& Std (M-gPC)  & 4.0931   & 2.5388   & 4.1026   & 3.5501   & 3.7875   & 4.0072   & 3.8193   & 4.1252   \\
& Std (MC)   & 3.7132   & 1.9014   & 3.7994   & 3.0927   & 3.3418   & 3.6469   & 3.4791   & 3.7686   \\
& RMSE       & 0.0125   & 0.0128   & 0.0124   & 0.0123   & 0.0125   & 0.0122   & 0.0122   & 0.0128   \\
& MAE        & 0.0097   & 0.0098   & 0.0096   & 0.0094   & 0.0096   & 0.0094   & 0.0094   & 0.01     \\
\bottomrule
\end{tabular} \normalsize
\end{table*}
\section{Conclusion}
\label{conclusion}
Uncertainty analysis for system design is an increasingly important concern especially when we navigate through the vast design space with many emerging and immature technologies.
The general MC method to analyze uncertainty is limited due to the expensive simulations in a practical architecture, while efficient modeling methods such as generalized polynomial chaos (gPC) cannot handle
the unique challenge when we move from the analog device world to a discrete architecture design. Many parameters and configurations are inherently discrete/integers at the system level, hence the uncertainty analysis becomes a mixed-domain problem.
To address these challenges, we propose a novel mixed-integer programming method to find a quadrature, and an M-gPC model that can handle both continuous and discrete inputs. 
The results of an analytical CMP model have shown that our framework with 95 samples can approximate the results of a MC method with $5\times {10^4}$ samples. 
We have also verified the proposed uncertainty analysis framework on detailed DRAM subsystems.
There are still many open problems for architectural uncertainty analysis, such as the high-dimensional cases and non-smooth output performance.
The proposed architectural uncertainty analysis framework can be used in many application cases in the future.

\section*{Acknowledgment}
This work was supported by NSF CCF No. 1763699 and NSF CAREER Award No. 1846476.

\appendices
\section{Basis Function Construction}
\label{sec:TTR}

When the uncertain parameters are mutually independent, a multivariate basis function can be constructed based on the product of some univariate ones:
\begin{equation}
{\Psi _{\basisInd}}\left( \vecpar \right) = \prod\limits_{i = 1}^d {\phi _{{\alpha _i}}^{\left( i \right)}} \left( {{\parm_i}} \right).
\end{equation}
Here $\phi_{\alpha_i}^{(i)}$ is a degree-$\alpha_i$ polynomial basis function for parameter $\parm_i$. Given the marginal probability density function $\rho_i\left( {{\parm_i}} \right)$ for each variable $\parm_i$, a set of uni-variate orthornormal polynomials ${\left\{ {\phi _m^{\left( i \right)},m \in \mathbb{N}} \right\}}$ can be constructed 
by the well-known three-term recurrence relation~\cite{gautschi1982generating}, which satisfy: 
\begin{equation}
    \mathbb{E}[\phi_{m}^{(i)}\phi_{n}^{(i)}]=\delta_{m,n}, \ \forall\, i=1,\ldots,d.
\end{equation}
Here ${\delta_{m,n}}$ is a Delta function. 
The basis functions for discrete variable are also defined in a discrete or integer domain.  

The three term recurrence~\cite{gautschi1982generating} is performed as follows:
\begin{equation}\label{eq:ttr}
\begin{array}{l}
{\pi _{i + 1}}\left( x \right) = \left( {x - {\alpha _i}} \right){\pi _i}\left( x \right) - {\beta _i}\left( x \right){\pi _{i - 1}}\left( x \right)\\
\begin{array}{*{20}{c}}
{{\pi _{ - 1}}\left( x \right) = 0,}&{{\pi _0}\left( x \right) = 1,}&{i = 0,1, \ldots ,n}
\end{array}
\end{array}
\end{equation}
where 
\begin{equation}\label{eq:three_term-coe}
\begin{array}{*{20}{c}}
{{\alpha _i} = \frac{{E\left[ {x\pi _i^2\left( x \right)} \right]}}{{E\left[ {\pi _i^2\left( x \right)} \right]}},}&{{\beta _{i + 1}} = \frac{{E\left[ {\pi _{i + 1}^2\left( x \right)} \right]}}{{E\left[ {\pi _i^2\left( x \right)} \right]}},}&{i = 0,1, \ldots ,n}
\end{array}
\end{equation}
and ${{\beta _{i}}} = 1$. 
Then the orthonormal polynomials are obtained via normalizing the above obtained polynomials: 
\begin{equation}
\label{eq:norm_poly}
\begin{array} {*{20}{c}}
{{\phi_i}\left( x \right) = \frac{{{\pi _i}\left( x \right)}}{{\sqrt {{\beta _0}{\beta _1} \cdots {\beta _i}} }},}&{i = 0,1, \ldots ,n}. 
\end{array}    
\end{equation}

\bibliographystyle{IEEEtran}
\small{
\bibliography{Bib/reference}

\begin{thebibliography}{10}
\providecommand{\url}[1]{#1}
\csname url@samestyle\endcsname
\providecommand{\newblock}{\relax}
\providecommand{\bibinfo}[2]{#2}
\providecommand{\BIBentrySTDinterwordspacing}{\spaceskip=0pt\relax}
\providecommand{\BIBentryALTinterwordstretchfactor}{4}
\providecommand{\BIBentryALTinterwordspacing}{\spaceskip=\fontdimen2\font plus
\BIBentryALTinterwordstretchfactor\fontdimen3\font minus
  \fontdimen4\font\relax}
\providecommand{\BIBforeignlanguage}[2]{{%
\expandafter\ifx\csname l@#1\endcsname\relax
\typeout{** WARNING: IEEEtran.bst: No hyphenation pattern has been}%
\typeout{** loaded for the language `#1'. Using the pattern for}%
\typeout{** the default language instead.}%
\else
\language=\csname l@#1\endcsname
\fi
#2}}
\providecommand{\BIBdecl}{\relax}
\BIBdecl

\bibitem{Mittal:2016:SAT:2891449.2871167}
S.~Mittal, ``A survey of architectural techniques for managing process
  variation,'' \emph{ACM Comput. Surv.}, vol.~48, no.~4, pp. 54:1--54:29, Feb.
  2016.

\bibitem{Mishra:2010:TCC:1773394.1773400}
A.~K. Mishra, J.~L. Hellerstein, W.~Cirne, and C.~R. Das, ``Towards
  characterizing cloud backend workloads: Insights from google compute
  clusters,'' \emph{SIGMETRICS Perform. Eval. Rev.}, vol.~37, no.~4, pp.
  34--41, Mar. 2010.

\bibitem{Borkar:2003:PVI:775832.775920}
S.~Borkar, T.~Karnik, S.~Narendra, J.~Tschanz, A.~Keshavarzi, and V.~De,
  ``Parameter variations and impact on circuits and microarchitecture,'' in
  \emph{Proc. DAC}, 2003, pp. 338--342.

\bibitem{Bhardwaj:2005:LMN:1065579.1065719}
S.~Bhardwaj and S.~B.~K. Vrudhula, ``Leakage minimization of nano-scale
  circuits in the presence of systematic and random variations,'' in
  \emph{Proc. Design Autom. Conf.}, 2005, pp. 541--546.

\bibitem{Zhang:2009:PVC:1629911.1630092}
L.~Zhang, L.~S. Bai, R.~P. Dick, L.~Shang, and R.~Joseph, ``Process variation
  characterization of chip-level multiprocessors,'' in \emph{Proc. Design
  Autom. Conf.}, 2009, pp. 694--697.

\bibitem{Wong:2005:FDA:1129601.1129608}
H.-Y. Wong, L.~Cheng, Y.~Lin, and L.~He, ``{FPGA} device and architecture
  evaluation considering process variations,'' in \emph{Proc. ICCAD}, 2005, pp.
  19--24.

\bibitem{das2007mitigating}
A.~Das, S.~Ozdemir, G.~Memik, J.~Zambreno, and A.~Choudhary, ``Mitigating the
  effects of process variations: Architectural approaches for improving batch
  performance,'' in \emph{Workshop on Arch. Support for Gigascale Integr.},
  2007.

\bibitem{7271059Yan}
G.~Yan, F.~Sun, H.~Li, and X.~Li, ``Corerank: Redeeming ``dark silicon" by
  dynamically quantifying core-level healthy condition,'' \emph{IEEE Trans.
  Computers}, vol.~65, no.~3, pp. 716--729, March 2016.

\bibitem{Cui:2017:EUA:3123939.3124541}
W.~Cui and T.~Sherwood, ``Estimating and understanding architectural risk,'' in
  \emph{Proc. MICRO}, 2017, pp. 651--664.

\bibitem{xiu2010numerical}
D.~Xiu, \emph{Numerical methods for stochastic computations: a spectral method
  approach}.\hskip 1em plus 0.5em minus 0.4em\relax Princeton university press,
  2010.

\bibitem{zhang2013stochastic}
Z.~Zhang, T.~A. El-Moselhy, I.~M. Elfadel, and L.~Daniel, ``Stochastic testing
  method for transistor-level uncertainty quantification based on generalized
  polynomial chaos,'' \emph{IEEE Trans. CAD Integr. Circuits Syst.}, vol.~32,
  no.~10, pp. 1533--1545, 2013.

\bibitem{zhang2013uncertainty}
Z.~Zhang, I.~A.~M. Elfadel, and L.~Daniel, ``Uncertainty quantification for
  integrated circuits: Stochastic spectral methods,'' in \emph{Proc. Int. Conf.
  Computer-Aided Design}, 2013, pp. 803--810.

\bibitem{zhang2016big}
Z.~Zhang, T.-W. Weng, and L.~Daniel, ``Big-data tensor recovery for
  high-dimensional uncertainty quantification of process variations,''
  \emph{IEEE Trans. Compon. Packag. Manuf. Techol.}, vol.~7, no.~5, pp.
  687--697, 2016.

\bibitem{cui2018uncertainty}
C.~Cui and Z.~Zhang, ``Uncertainty quantification of electronic and photonic
  {ICs} with non-{Gaussian} correlated process variations,'' in \emph{Proc.
  ICCAD}, 2018, pp. 1--8.

\bibitem{agarwal2009domain}
N.~Agarwal and N.~R. Aluru, ``A domain adaptive stochastic collocation approach
  for analysis of mems under uncertainties,'' \emph{J. Comp. Phys.}, vol. 228,
  no.~20, pp. 7662--7688, 2009.

\bibitem{xiu2005high}
D.~Xiu and J.~S. Hesthaven, ``High-order collocation methods for differential
  equations with random inputs,'' \emph{SIAM Journal on Scientific Computing},
  vol.~27, no.~3, pp. 1118--1139, 2005.

\bibitem{sarangi2008varius}
S.~R. Sarangi, B.~Greskamp, R.~Teodorescu, J.~Nakano, A.~Tiwari, and
  J.~Torrellas, ``Varius: A model of process variation and resulting timing
  errors for microarchitects,'' \emph{IEEE Trans. Semiconductor Manufacturing},
  vol.~21, no.~1, pp. 3--13, 2008.

\bibitem{chandrasekar2014exploiting}
K.~Chandrasekar, S.~Goossens, C.~Weis, M.~Koedam, B.~Akesson, N.~Wehn, and
  K.~Goossens, ``Exploiting expendable process-margins in {DRAMs} for run-time
  performance optimization,'' in \emph{Proc. Design, Autom. \& Test in Europe},
  2014, p. 173.

\bibitem{rosenfeld2011dramsim2}
P.~Rosenfeld, E.~Cooper-Balis, and B.~Jacob, ``{DRAMSim2}: A cycle accurate
  memory system simulator,'' \emph{IEEE Computer Architecture Letters},
  vol.~10, no.~1, pp. 16--19, 2011.

\bibitem{cui2018stochastic}
C.~Cui, M.~Gershman, and Z.~Zhang, ``Stochastic collocation with non-{Gaussian}
  correlated parameters via a new quadrature rule,'' in \emph{Proc. IEEE Conf.
  EPEPS}, 2018, pp. 57--59.

\bibitem{cui2018stochastic_journal}
C.~Cui and Z.~Zhang, ``Stochastic collocation with non-{Gaussian} correlated
  process variations: Theory, algorithms, and applications,'' \emph{IEEE Trans.
  Compon. Packag. Manuf. Techol.}, vol.~9, no.~7, pp. 1362--1375, July 2019.

\bibitem{grossmann2002review}
I.~E. Grossmann, ``Review of nonlinear mixed-integer and disjunctive
  programming techniques,'' \emph{Optimization and engineering}, vol.~3, no.~3,
  pp. 227--252, 2002.

\bibitem{hansen1992analysis}
P.~C. Hansen, ``Analysis of discrete ill-posed problems by means of the
  l-curve,'' \emph{SIAM review}, vol.~34, no.~4, pp. 561--580, 1992.

\bibitem{hill2008amdahl}
M.~Hill and M.~Marty, ``Amdahl's law in the multicore era,'' \emph{Computer},
  vol.~41, no.~7, pp. 33--38, 2008.

\bibitem{spec2017}
``Spec cpu 2017,'' \url{https://www.spec.org/cpu2017/}, accessed: 2019-04-08.

\bibitem{gautschi1982generating}
W.~Gautschi, ``On generating orthogonal polynomials,'' \emph{SIAM J. Sci. Stat.
  Comp.}, vol.~3, no.~3, pp. 289--317, 1982.

\end{thebibliography}
}

\end{document}